\documentclass[twocolumn,english,aps,prl,showpacs,superscriptaddress]{revtex4-1}
\usepackage[T1]{fontenc}
\usepackage[latin9]{inputenc}
\usepackage[dvipsnames]{xcolor}
\usepackage{units}
\usepackage{amsmath}
\usepackage{amssymb}
\usepackage{graphicx}
\usepackage{braket}

\makeatletter

\newcommand*\LyXThinSpace{\,\hspace{0pt}}
\providecommand{\tabularnewline}{\\}
\IfFileExists{lmodern.sty}{\usepackage{lmodern}}{}
\usepackage{babel}
\usepackage{dsfont}

\usepackage[bookmarks=false]{hyperref}
\def\equationautorefname~#1\null{Equation (#1)\null}

\newcommand{\BL}[1]{{\color{black}#1}}

\makeatother

\usepackage{babel}
\begin{document}
\title{{Optical quantum memory for noble-gas spins based on spin-exchange collisions}}
\author{Or Katz}
\thanks{These authors contributed equally to this work.}
\affiliation{Department of Physics of Complex Systems, Weizmann Institute of Science,
Rehovot 76100, Israel}
\affiliation{Rafael Ltd, IL-31021 Haifa, Israel}
\address{present address: Department of Electrical and Computer Engineering, Duke University, Durham, North Carolina 27708, USA}
\email{or.katz@duke.edu}
\author{Roy Shaham}
\thanks{These authors contributed equally to this work.}
\affiliation{Department of Physics of Complex Systems, Weizmann Institute of Science,
Rehovot 76100, Israel}
\affiliation{Rafael Ltd, IL-31021 Haifa, Israel}
\author{Eran Reches}
\affiliation{Department of Physics of Complex Systems, Weizmann Institute of Science,
Rehovot 76100, Israel}
\author{Alexey V. Gorshkov}
\affiliation{Joint Quantum Institute and Joint Center for Quantum Information and
Computer Science, NIST/University of Maryland, College Park, Maryland
20742, USA}
\author{Ofer Firstenberg}
\affiliation{Department of Physics of Complex Systems, Weizmann Institute of Science,
Rehovot 76100, Israel}
\begin{abstract}
{Optical quantum memories, which store and preserve the quantum state of photons, rely on a coherent mapping of the photonic state onto matter states that are optically accessible. Here we outline and characterize schemes to map the state of photons onto long-lived but optically-inaccessible collective states of noble-gas spins. The mapping employs coherent spin-exchange interaction arising from random collisions with alkali vapor. We propose efficient storage strategies in two operating regimes and analyze their performance for several proposed  experimental configurations.} 
\end{abstract}
\maketitle

\section{Introduction}

Optical quantum memories enable the storage and retrieval of an optical
signal while preserving its quantum properties. In quantum
information science, they can function as variable delay lines,
leaving a quantum state unchanged during the memory operation \citep{Tittel-review-nature-photonics}.
Optical quantum memories are vital for applications such as quantum
repeaters in long-distance quantum communication \citep{Gisin-repeaters,DLCZ,zellinger,Heshami},
conversion of heralded photons to on-demand single photons \citep{de-Riedmatten-rare-earth-PRL-2019,Wang-2019,Polzik-single-photon},
synchronization of optical quantum computation \citep{oPTICAL-QUANTUM-COMPUTING-SCIENCE},
distribution of entanglement \citep{Kimble-2008,entang1}, and metrology beyond
the standard quantum limit \citep{Polzik-2011,Lvovsky-2008,Yudin-2010,Kominis-2008,Kong-2020}.

Existing quantum memories are based on actual delay lines and cavities
\citep{Tittel-review-nature-photonics} or on a reversible mapping
onto coherent excitations in matter \citep{Polzik-RMP-2010}. The
latter relies on strong light-matter coupling, known as the cooperativity
(or the collective cooperativity in ensembles) which determines the memory efficiency \citep{Gorshkov-PRL,Gorshkov1,Gorshkov2}. The mapping process
employs a variety of schemes, including electromagnetically induced
transparency (EIT) \citep{Novikova-review,Lukin-2003,Fleischhauer-EIT,Duan-2020},
off-resonant Raman interaction \citep{Raman1,Raman2,Raman3}, teleportation
via the Faraday interaction \citep{Polzik-coherent-state-memory,Polzik-2002,Pinard-teleportation},
and a range of echo techniques \citep{Buchler-GEM,CRIB}. Different media can be utilized for the memory, including solid media,
such as defects in diamonds and rare-earth-doped crystals \citep{Lukin-SIV,Halfmann,Serrano-rare-earth},
as well as gaseous medium, primarily cold or warm atomic alkali-metal atoms
\citep{Lukin-PRL-2001,Eisaman-2005,Katz-storage-of-light-2018,Goldfarb-CPO,Riedmatten-rydberg-memory,Kuzmich-2013}. 

{The length of time for which the memory can store the information (memory lifetime) is governed by the isolation of the matter excitation from the environment. Crystals doped with rare-earth ions have demonstrated lifetimes up to a millisecond for storage of non-classical light \citep{EuMiliSecQuantum_PRL2017,RiedmattenSolidStateShortStorage,TittelSolideStateShortStorage} and up to an hour for storage of coherent light, with sub-percent efficiency and at cryogenic temperatures \citep{Halfmann,EuOneHourClassic_NatureComm2021}. As some ions feature direct optical access and hours-long coherence times \citep{EuSixHoursClassic_Nature2015,EuOneHourClassic_NatureComm2021}, they can potentially be utilized as quantum memories with hours-long lifetimes once their efficiency is improved. 
}

{Optical memories based on atomic spin gases are another prominent candidate for optical quantum memories. They use the collective states of alkali spin ensembles, which feature collective strong coupling to light and strong resilience to local depolarization of individual atoms \citep{Thomas_FWM_2019,Michelberger_FWM_2015,Polzik-coherent-state-memory,Polzik-2011,Romalis-stroboscopic-2011,Vasilakis_Spin_Noise_2011,Krauter_Spin_Noise_2011,Molmer_Spin_States_2002}. This technology has demonstrated storage, generation, and retrieval of non-classical light despite thermal motion \citep{LukinBuffer_2003,Polzik-coherent-state-memory,LukinPulseShaping_2004,Eisaman-2005,Polzik-2011,LukinPulseShaping_2004,Borregaard_2016,Polzik-single-photon,Sun_2019,Bao_2020,Kong-2020,entang1,Li_arxiv_2020,PolzikMilisecRoomTempMemory}, imperfect initial polarization \citep{Polzik-coherent-state-memory,Polzik-2011}, and collisions of the alkali atoms with buffer gas \citep{Eisaman-2005,LukinBuffer_2003,entang1}. These experiments are commonly described by the Heisenberg-Bloch-Langevin model, which captures the irreversible nature of the relaxation processes and quantifies the effect of noise on the quantum memory \citep{Thomas_FWM_2019,Romalis-stroboscopic-2011,Vasilakis_Spin_Noise_2011,Krauter_Spin_Noise_2011,CollectiveSpinStatesThermalDynamics,Firstenberg-Weak-collisions}. Yet, the lifetime of alkali-based memories is limited by the coupling of the valence electron's spin to the environment, which typically sets the lifetime in the range of microseconds to hundreds of milliseconds \citep{Buchler-efficiency-storageTime-review,Katz-storage-of-light-2018}, 
except for special configurations that potentially enable efficient storage for up to tens of seconds \citep{Kuzmich-2013,Budker-1-MIN}.}

Rare isotopes of noble gases, such as $^{3}$He,
possess nonzero nuclear spins, which are isolated from the environment
due to the enclosing, complete, electronic shells. These spins exhibit
hours to days long lifetimes at or above room temperature \citep{Walker-RMP-2017,Gemmel-60-hours-coherence-time-He-2010,PolarizedHeliumCellsNIST} and are employed for medical lung imaging \citep{MRI1,MRI2}, for
precision magnetometers and NMR \citep{Chupp-noble-gas-magnetometer,romalis-2005,Walker2006,Walker-2019},
for neutron spin filters \citep{Gentile-2005}, and for searches of
beyond-standard-model physics \citep{PM1,PM2,PM3,PM4,PM5,PM6}. Unfortunately,
these spins are optically inaccessible and are thus extremely hard
to prepare, interface, and monitor. They are accessible, however,
through collisions with other optically-active atoms, either metastable
helium atoms via \emph{strong} metastable-exchange collisions \citep{Walker-RMP-2017,MEOP}
or alkali-metal atoms via \emph{weak} spin-exchange collisions {\citep{Walker-RMP-2017,Happer-Walker-RMP,Happer-1998,HapperSEOPEfficiency,BouchiatNuclearPolarization}}.

The first proposal to utilize noble-gas spins as a quantum memory
was based on the former, with metastable helium population sustained
by electric discharge \citep{Sinatra-HE3-memory-2005}. Metastable
exchange collisions rely on the strong electrostatic exchange interaction,
leading to a complete transfer of the electronic configuration in a
single collision (hence the terminology ``strong'' collision). The ground-state atom may be excited to the metastable state, and the metastable atom is then de-excited to the electronic ground-state, but both atoms maintain their nuclear states, and thus these collisions act to equilibrate the nuclear spin between the metastable and ground-state populations. A memory scheme relying on metastable exchange necessitates high collective optical cooperativity of the metastable population, which in turn requires either high-finesse cavities or higher helium
densities (practically limited by Penning collisions), and the scheme has never been demonstrated.

Spin-exchange collisions between alkali and noble-gas atoms, on the
other hand, involve the weak isotropic hyperfine interaction, where
only a small fraction ($10^{-4}-10^{-6}$) of the spin orientation
is transferred in a single collision \citep{Happer-Walker-RMP}. {Numerous
collisions can then accumulate to a collective, coherent evolution
of the two spin ensembles, leading to an exchange of collective excitations
that is free from excess thermalization and quantum noise \citep{Firstenberg-Weak-collisions}.}
{Unlike with metastable helium, alkali vapor density (determined by the temperature of the enclosing cell) is independent of noble-gas density (determined
by pressure). It is therefore feasible to increase alkali density
and reach high optical cooperativity. At the same time, the coherent
coupling between the collective alkali and collective noble-gas spins can
be made efficient and controllable with external fields, as demonstrated in Refs.~\cite{shaham-strong-coupling,Kornack2002}.} {A coherent bi-direcitonal coupling between light and noble-gas spins was demonstrated in Ref.~\cite{katz-spectro} in a continuous spectroscopic measurement, and this coupling was also proposed to generate two-mode spin squeezed states between two distant noble-gas cells \citep{Firstenberg-QND}. However, reversible, time dependent mapping
of non-classical optical signals onto noble-gas spins using the spin-exchange interaction has never been studied.}

In this paper, we propose a mechanism for mapping the quantum state of photons onto the macroscopic spin state of noble gases via spin-exchange collisions. We derive the equations of motion of the system in a Bloch-Heisenberg-Langevin framework. The derivation includes various factors affecting the memory performance, such as the atomic thermal motion, nonuniform spatial profile of the optical
fields, possible spin relaxations at the cell walls, and noise induced by imperfect polarization of the spin ensembles or by incoherent excitations. Subsequently,
we present two exemplary storage protocols that result in efficient mapping for low- and high-bandwidth optical pulses, respectively. We investigate several feasible experimental conditions for realizing such memories, including various alkali-metal and noble-gas
mixtures, and a range of temperatures, coatings, and gas pressures.
This work can thus be used to design and realize viable quantum memories
using the macroscopic quantum states of
noble-gas spins.

\BL{\section{Overview of main results}
In Sec.~\ref{sec:model}, we present the system under study and derive Eqs.~(\ref{eq:P_r_t_equation})-(\ref{eq:Input_Output_relation}). These equations describe the coupling of the input and output signal light fields to the alkali spins, based on the dipole interaction and the model of Ref.~\citep{Gorshkov1}, as well as the local coherent spin-exchange interaction between the alkali and noble-gas spins, based on Ref.~\citep{Firstenberg-Weak-collisions}. Owing to thermal motion of the atoms, the dynamics is best represented by non-local collective spin modes. In Sec.~\ref{sec:spatial_modes}, we simplify our analysis and focus on the dynamics of the least decaying (non-local) spin modes, which is described by Eqs.~(\ref{eq:output-light-P})-(\ref{eq:Uniform_alkali_spin_mode_diff_equation}) and (\ref{eq:Uniform_noble_spin_mode_diff_equation}). The full multimode analysis is detailed in Appendix \ref{sec:Spatial-modes-representation}.

In Sec.~\ref{sec:mem_eff}, we define the memory efficiency as a metric for the system performance. In Sec.~\ref{sec:Numerical-storage}, we present and exemplify two different memory protocols featuring high memory efficiencies [cf.~Eqs.(\ref{eq:efficiency of the fast storage scheme})-(\ref{eq:memory efficiency adiabatic})]. The first protocol relies on efficient exchange between the alkali and noble-gas spins in the strong-coupling regime \citep{shaham-strong-coupling}, which enables high-bandwidth storage ($>1$ MHz) limited only by the excited-state dynamics of the alkali spins. The second protocol alleviates the requirement for strong coupling between the gases on the expanse of lower memory bandwidth. 

In Sec.~\ref{sec:Stochastic-evolution}, we outline and discuss various experimental configurations which can potentially realize the proposed protocols, exploring different alkali vapor and noble-gas mixtures, different cell coatings, and different protocols for realizing $\Lambda$-type coupling which is compatible with the spin-exchange interaction, as summarized in table \ref{tab:diffusion-decay-rates}. In Sec.~{\ref{sec:Noise_characterization}}, we discuss the effect of fluorescence-noise as well as imperfect polarization of the spin ensembles on the memory performance, and finally summarize our work in Sec.~\ref{sec:Summary}.}

\section{Model of the system}\label{sec:model}
\subsection{System constituents}

\begin{figure}[t]
\begin{centering}
\includegraphics[clip,width=7.5cm]{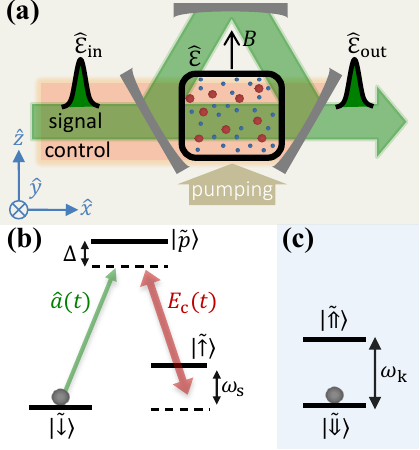}
\par\end{centering}
\centering{}\caption{\textbf{(a)} A quantum memory system based on a mixture of noble-gas
and alkali spins. The alkali-metal atoms (red) and noble-gas atoms
(blue) interact with the quantum signal field $\hat{\mathcal{E}}$
in a cavity and with a classical control field. The atomic spins are initially polarized with an auxiliary pumping beam. The signal
field in the cavity is coupled to the incoming field $\hat{\mathcal{E}}_{\text{in}}$,
to be stored, and to the output field $\hat{\mathcal{E}}_{\text{out}}$,
retrieved on demand after the memory time. \textbf{(b)} Energy levels
of the modeled alkali atom in the lab frame.  The $\Lambda$ system
consists of two stable ground-level states $|\tilde{\uparrow}\rangle$
and $|\tilde{\downarrow}\rangle$, coupled via an excited state $|\tilde{\mathrm{p}}\rangle$
by the classical control field $E_{\textrm{c}}$ and the quantum signal $\hat{a}=e^{-i\omega_{\varepsilon}t}\hat{\mathcal{E}}$.
Initially, the state $|\tilde{\downarrow}\rangle$ is populated.
\textbf{(c)} Energy levels of the modeled noble-gas atom in the lab
frame. Initially, the state $|\tilde{\Downarrow}\rangle$ is populated.
\label{fig: setup}}
\end{figure}

Consider a glass cell of volume $V$ containing $N_{\text{a}}$
alkali-metal spins and $N_{\text{b}}$ noble-gas spins, positioned
inside an optical cavity as shown in Fig.~\ref{fig: setup}(a). The
alkali spins are initially polarized along the $\hat{z}$ axis by
optical pumping, and the noble-gas spins are hyper-polarized via spin-exchange
optical pumping (SEOP) \citep{Happer-1998,Happer-Book,tsinovoy}. In the presence
of an optical quantum field, which serves as the signal, the total
Hamiltonian of the system is given by
\begin{equation}
\tilde{\mathcal{H}}=\tilde{\mathcal{H}}_{\varepsilon}+\tilde{\mathcal{H}}_{\text{a}}+\tilde{\mathcal{H}}_{\text{b}}+\tilde{\mathcal{H}}_{\text{a-}\varepsilon}+\tilde{\mathcal{H}}_{\text{a-b}}.
\end{equation}
Here $\tilde{\mathcal{H}}_{\varepsilon}$ is the Hamiltonian of the
signal light field, $\tilde{\mathcal{H}}_{\text{a}}$ is the Hamiltonian
of the alkali-metal spins, $\tilde{\mathcal{H}}_{\text{b}}$ is the
Hamiltonian of the nuclear spins of the noble gas, $\tilde{\mathcal{H}}_{\text{a-}\varepsilon}$
is the atom-light dipole interaction, and $\tilde{\mathcal{H}}_{\text{a-b}}$
is the coherent spin-exchange interaction between the electronic spins
of the alkali atoms and the nuclear spins of the noble-gas atoms \citep{Firstenberg-Weak-collisions}.

Each polarized alkali-metal atom, labeled by $a$, is modeled as a
three-level system in a $\Lambda$ configuration, consisting of two
ground-level states $|\tilde{\downarrow}\rangle_{a}$ and $|\tilde{\uparrow}\rangle_{a}$
and a single excited state $\left|\tilde{\mathrm{p}}\right\rangle _{a}$,
as shown in Fig.~\ref{fig: setup}(b). The corresponding Hamiltonian
is
\begin{equation}
\tilde{\mathcal{H}}_{\text{a}}=\hbar\sum_{a=1}^{N_{\mathrm{a}}}\left(\omega_{\textnormal{p}}\left|\tilde{\mathrm{p}}\right\rangle _{a}\left\langle \tilde{\mathrm{p}}\right|_{a}+\omega_{\textnormal{s}}|\tilde{\uparrow}\rangle_{a}\langle\tilde{\uparrow}|_{a}\right),
\end{equation}
where $\omega_{\textnormal{p}}$ is the resonance frequency of the
optical transition $|\tilde{\downarrow}\rangle-\left|\tilde{\mathrm{p}}\right\rangle $, and $\omega_{\textnormal{s}}$ is the frequency difference
between $|\tilde{\downarrow}\rangle$ and $|\tilde{\uparrow}\rangle$.
If the two spin states are in the same hyperfine manifold, then $\omega_{\textnormal{s}}=g_{\textnormal{s}}B$
corresponds to the Zeeman splitting, where $\mathbf{B}=B\hat{z}$
is the magnetic field, and $g_{\textnormal{s}}=2.8/\left[I\right]\times2\pi\text{\,MHz/G}$
is the gyromagnetic ratio of an alkali atom with nuclear spin $I$
($\left[I\right]\equiv2I+1$).

We consider noble-gas atoms with nuclear spin-1/2 (\textit{e.g}.,~$^{3}\text{He}$
and $^{129}\text{Xe}$). Each noble-gas atom, labeled by
$b$, consists of two spin levels $|\tilde{\Downarrow}\rangle_{b}$
and $|\tilde{\Uparrow}\rangle_{b}$, as shown in Fig.~\ref{fig: setup}(c).
The corresponding Hamiltonian is
\begin{equation}
\tilde{\mathcal{H}}_{\text{b}}=\hbar\sum_{b=1}^{N_{\text{b}}}\omega_{\mathrm{k}}|\tilde{\Uparrow}\rangle_{b}\langle\tilde{\Uparrow}|_{b},
\end{equation}
with $\omega_{\text{k}}=g_{\text{k}}B$ being the frequency difference
between the states $|\tilde{\Uparrow}\rangle$ and $|\tilde{\Downarrow}\rangle$
due to the Zeeman splitting of the noble-gas spins with gyromagnetic ratio
$g_{\text{k}}$.

We shall adopt a cavity model for describing the quantum optical field
\citep{Gorshkov1}. We assume that the signal field resides in a single
mode of a running-wave cavity, as described by the Hamiltonian
\begin{equation}
\tilde{\mathcal{H}}_{\varepsilon}=\hbar\omega_{\varepsilon}\hat{a}^{\dagger}\hat{a}.
\end{equation}
Here $\hat{a}$ is the bosonic annihilation operator of the electromagnetic
field of the cavity mode with frequency $\omega_{\varepsilon}$.

The atom-photon interaction Hamiltonian in the dipole approximation is given by
\begin{equation}
\tilde{\mathcal{H}}_{\text{a-}\varepsilon}=-\sum_{a=1}^{N_{a}}\hat{\text{\textbf{d}}}_{a}\cdot\hat{\boldsymbol{E}}(\mathbf{r}_{a},t),
\end{equation}
where
$\hat{\text{\textbf{d}}}_{a}$
is the dipole operator of the alkali-metal atoms, and $\boldsymbol{\hat{\boldsymbol{E}}}(\mathbf{r}_{a},t)$
is the electric field at the location $\mathbf{r}_{a}$ of that atom.
In our model, the electric field is composed of a classical control field and a quantum signal. The control field, with frequency $\omega_{\mathrm{c}}$ and amplitude $E_{\mathrm{c}}f_{\text{c}}(\mathbf{r})$, couples to the $|\tilde{\downarrow}\rangle_{a} - |\tilde{\mathrm{p}}\rangle_{a}$ transition, with dipole moment $\mu_\text{c}$. The quantum signal field $\sqrt{\hbar\omega_{\varepsilon}/(2\epsilon_{0})}f_{\varepsilon}\left(\mathbf{r}\right)\hat{a}(t)$ couples to the $|\tilde{\uparrow}\rangle_{a} - |\tilde{\mathrm{p}}\rangle_{a}$ transition, with dipole moment $\mu_\varepsilon$. We thus obtain
\begin{align}
\tilde{\mathcal{H}}_{\text{a-}\varepsilon}=&-\sum_{a=1}^{N_{a}} f_\text{c}(\mathbf{r}_a)
\mu_\text{c} E_{\text{c}}(t) e^{-i\omega_{\text{c}}t}  |\tilde{\uparrow}\rangle_a\langle\tilde{\mathrm{p}}|_a+\text{h.c.} \nonumber \\
 &-\sum_{a=1}^{N_{a}} f_\varepsilon(\mathbf{r}_a) 
 \mu_\varepsilon   \sqrt{\frac{\hbar\omega_{\varepsilon}}{2\epsilon_{0}}}\hat{a}(t)
|\tilde{\downarrow}\rangle_a\langle\tilde{\mathrm{p}}|_a+\text{h.c.}
\end{align}

The spatial mode functions 
$f_{\text{c}}\left(\mathbf{r}\right)$ and $f_{\varepsilon}\left(\mathbf{r}\right)$ satisfy the Helmholtz equation
\begin{equation}
\left(\nabla^{2}+k_{i}^{2}\right)f_{i}\left(\mathbf{r}\right)=0,\label{eq:Helmholtz equation}
\end{equation}
where $i\in\left\{\varepsilon,\text{c}\right\} $, with the boundary
conditions determined by the cavity: $f_{\text{c}}\left(\mathbf{r}\right)$
is the solution with an eigenvalue $k_{\text{c}}=\omega_{\text{c}}/c$,
and $f_{\varepsilon}\left(\mathbf{r}\right)$ is the solution with
an eigenvalue $k_{\varepsilon}=\omega_{\varepsilon}/c$, where $c$
denotes the speed of light. The Rabi frequency
of the classical field within the cavity is given by $\Omega\left(\mathbf{r},t\right)=\sqrt{V_{\text{cav}}}f_{\text{c}}\left(\mathbf{r}\right)\Omega\left(t\right)$,
where $\Omega\left(t\right)=\mu_{\text{c}}E_{\text{c}}\left(t\right)/(\hbar\sqrt{V_{\text{cav}}}$).
Similarly, $g\left(\mathbf{r}\right)=\sqrt{V_{\text{cav}}}gf_{\varepsilon}\left(\mathbf{r}\right)$
is the one-photon Rabi frequency for the quantized-field mode, where \citep{Gorshkov1}
\begin{equation}
g=\mu_{\varepsilon}\sqrt{\omega_{\varepsilon}/(2\epsilon_{0}\hbar V_{\text{cav}})}.\label{eq:definition of G}
\end{equation}

Before discussing the spin-exchange interaction $\tilde{\mathcal{H}}_{\text{a-b}}$,
we transform the above Hamiltonians into a rotating
frame.  The transformation is given by
\begin{equation}
U_{a}=e^{i\omega_{\varepsilon}t}|\mathrm{p}\rangle_{a}\langle\tilde{\mathrm{p}}|_{a}+e^{i(\omega_{\varepsilon}-\omega_{\textnormal{c}})t}\left|\uparrow\right\rangle _{a}\langle\tilde{\uparrow}|_{a}+\left|\downarrow\right\rangle _{a}\langle\tilde{\downarrow}|_{a}
\end{equation}
for any alkali spin $a$, and by
\begin{equation}
U_{b}=e^{i(\omega_{\varepsilon}-\omega_{\textnormal{c}})t}\left|\Uparrow\right\rangle _{b}\langle\tilde{\Uparrow}|_{b}+\left|\Downarrow\right\rangle _{b}\langle\tilde{\Downarrow}|_{b}
\end{equation}
for any noble-gas spin $b$. We also use the transformation  $U_{\varepsilon}=e^{i\tfrac{t}{\hbar}\mathcal{H}_{\varepsilon}}$ for the signal field and define the slowly varying quantum field operator  $\hat{\mathcal{E}}\left(t\right)=e^{i\omega_{\varepsilon}t}\hat{a}\left(t\right)$,
which describes the envelope of the quantum field within the cavity. We further define the slowly-varying continuous atomic operators
\begin{align}
\hat{\sigma}_{\mu\nu}\left(\mathbf{r},t\right)=\sum_{a=1}^{N_{\mathrm{a}}}\left|\mu\right\rangle _{a}\left\langle \nu\right|_{a}\delta(\mathbf{r}-\mathbf{r}_{a}) & ,
\end{align}
which describe the collective state of the alkali ensemble with $\mu,\nu\in\left\{ \downarrow,\uparrow,\textnormal{p}\right\} $,
and 
\begin{equation}
\hat{\sigma}_{\mu\nu}\left(\mathbf{r},t\right)=\sum_{b=1}^{N_{\mathrm{b}}}\left|\mu\right\rangle _{b}\left\langle \nu\right|_{b}\delta(\mathbf{r}-\mathbf{r}_{b}),
\end{equation}
which describe the collective state of the noble-gas ensemble with $\mu,\nu\in\left\{ \Downarrow,\Uparrow\right\} $. In the rotating
frame, we get
\begin{align}
\mathcal{H-H_{\text{a-b}}=} & \hbar\int_{V}d^{3}\mathbf{r}\,\Delta\hat{\sigma}_{\textnormal{pp}}+\tilde{\delta}_{\textnormal{s}}\hat{\sigma}_{\uparrow\uparrow}+\tilde{\delta}_{\textnormal{k}}\hat{\sigma}_{\Uparrow\Uparrow}\label{eq:Hamiltonian_tot}\\
- & \Bigl[\Omega\left(\mathbf{r},t\right)\hat{\sigma}_{\textnormal{p}\uparrow}+g\left(\mathbf{r}\right)\hat{\mathcal{E}}(t)\hat{\sigma}_{\textnormal{p}\downarrow}+\text{h.c.}\Bigr],\nonumber
\end{align}
where $\Delta=\omega_{\textnormal{p}}-\omega_{\varepsilon}$ is the
one-photon detuning from the atomic optical transition, and $\tilde{\delta}_{\textnormal{s}}=\omega_{\textnormal{s}}+\omega_{\text{c}}-\omega_{\varepsilon}$
and $\tilde{\delta}_{\textnormal{k}}=\omega_{\textnormal{k}}+\omega_{\text{c}}-\omega_{\varepsilon}$
are the two-photon detunings from the spin resonances of the two species (absent the shifts induced by the spin-exchange
interaction).

\subsection{Spin-exchange coupling}

The two spin gases experience random, weak, spin-exchange collisions.
For polarized gases, the leading term in the dynamics is described
by the coherent interaction Hamiltonian \citep{Firstenberg-Weak-collisions}
\begin{equation}
\tilde{\mathcal{H}}_{\text{a-b}}=\hbar \zeta\int\mathrm{d}^{3}\mathbf{r}_{1}\int\mathrm{d}^{3}\mathbf{r}_{2}\delta(\mathbf{r}_{1}-\mathbf{r}_{2})\boldsymbol{\hat{\textbf{f}}}(\mathbf{r}_{1},t)\cdot\boldsymbol{\hat{\textbf{k}}}(\mathbf{r}_{2},t),\label{eq:a-b hamiltonian}
\end{equation}
where $\boldsymbol{\hat{\textbf{f}}}\left(\mathbf{r},t\right)\equiv\sum_{a}\boldsymbol{\hat{\textbf{f}}}_{a}\left(t\right)\delta\bigl(\mathbf{r}-\mathbf{r}_{a}\left(t\right)\bigr)$
and $\boldsymbol{\hat{\textbf{k}}}\left(\mathbf{r},t\right)\equiv\sum_{b}\boldsymbol{\hat{\textbf{k}}}_{b}\left(t\right)\delta\bigl(\mathbf{r}-\mathbf{r}_{b}\left(t\right)\bigr)$
denote the continuous spin operators of alkali and noble-gas spins,
respectively. $\zeta\delta(\mathbf{r}_{1}-\mathbf{r}_{2})$ is the local
average interaction strength of an alkali and noble-gas atom pair,
where the microscopic interaction strength constant $\zeta=\left\langle \phi\sigma v\right\rangle _{\text{c}}/\left[I\right]$
is given by ensemble averaging over all realizations of the collisional
parameters, given the velocity $v$, the hard-sphere cross-section
$\sigma$, and the accumulated phase $\phi$ during a single collision
instance.

The ground level of actual alkali-metal atoms consists of multiple
spin levels. We choose $\mid\downarrow\rangle_{a}$ to be the maximally-polarized
spin state (with the projection $I+1/2$ along the quantization axis)
and choose $\mid\uparrow\rangle_{a}$ to be the adjacent state (with
projection $I-1/2$); both states are in the hyperfine manifold
$F=I+1/2$. With this choice and for fully polarized ensembles, it
is a good approximation to replace the total spin operator $\boldsymbol{\hat{\mathrm{f}}}$
by its projection on the two-state subsystem $\boldsymbol{\hat{\textbf{f}}}\approx\hat{P}\boldsymbol{\hat{\textbf{f}}}\hat{P}$,
where $\hat{P}=\sum_{a}\left(\left|\uparrow\right\rangle _{a}\left\langle \uparrow\right|_{a}+\left|\downarrow\right\rangle _{a}\left\langle \downarrow\right|_{a}\right)$.
We then obtain in the rotating frame
\begin{equation}
\begin{aligned}\boldsymbol{\hat{\textbf{f}}}\left(\mathbf{r},t\right) & \approx\tfrac{\left[I\right]}{2}\bigl(\hat{\sigma}_{\downarrow\downarrow}+q_{I}\hat{\sigma}_{\uparrow\uparrow}\bigr)\boldsymbol{e}_{z}\\
+ & \sqrt{\tfrac{[I]}{2}}\bigl(e^{i(\omega_{\text{c}}-\omega_{\varepsilon})t}\hat{\sigma}_{\downarrow\uparrow}\boldsymbol{e}_{-}+\text{h.c.}\bigr),
\end{aligned}
\label{eq:f_spin}
\end{equation}
where $\boldsymbol{e}_{\pm}=(\boldsymbol{e}_{x}\pm i\boldsymbol{e}_{y})/\sqrt{2}$
and $q_{I}=\left(\left[I\right]-2\right)/\left[I\right]$. 

The collective
noble-gas spin operator appearing in Eq.~(\ref{eq:a-b hamiltonian})
is given in the rotating frame by
\begin{equation}
\begin{aligned}\boldsymbol{\hat{\textbf{k}}}\left(\mathbf{r},t\right) & =\tfrac{1}{2}\bigl(\hat{\sigma}_{\Downarrow\Downarrow}-\hat{\sigma}_{\Uparrow\Uparrow}\bigr)\boldsymbol{e}_{z}\\
+ & \sqrt{\tfrac{1}{2}}\bigl(e^{i(\omega_{\text{c}}-\omega_{\varepsilon})t}\hat{\sigma}_{\Downarrow\Uparrow}\boldsymbol{e}_{-}+\text{h.c.}\bigr).
\end{aligned}
\label{eq:Bloch-dynamics}
\end{equation}
We thus arrive at the spin-exchange Hamiltonian
\begin{align}
\mathcal{H}_{\mathrm{a-b}} & =\hbar\int_{V}d^{3}\mathbf{r}\left[\mathcal{H}_{\text{s}}+\zeta\sqrt{\left[I\right]}\bigl(\hat{\sigma}_{\downarrow\uparrow}\hat{\sigma}_{\Uparrow\Downarrow}+\text{h.c.}\bigr)/2\right],\label{eq:Hamiltonian_tot-1}
\end{align}
where the first term
\begin{equation}
\mathcal{H}_{\text{s}}(\mathbf{r},t)=\zeta\left[I\right]\bigl(\hat{\sigma}_{\downarrow\downarrow}+q_{I}\hat{\sigma}_{\uparrow\uparrow}\bigr)\bigl(\hat{\sigma}_{\Downarrow\Downarrow}-\hat{\sigma}_{\Uparrow\Uparrow}\bigr)/4
\end{equation}
describes an additional energy shift, which becomes prominent for
polarized ensembles. The second term in Eq.~(\ref{eq:Hamiltonian_tot-1})
manifests the conservative exchange of spin between the two gases.

\subsection{Dissipation and atomic motion}

The system Hamiltonian $\mathcal{H}$ given by Eqs.~(\ref{eq:Hamiltonian_tot})
and (\ref{eq:Hamiltonian_tot-1}) constitutes the unitary evolution
of the system. The system is however coupled to the environment: the
cavity field decays at a rate $\kappa$ through the output
port; the optical coherence between $\ket{\downarrow}$ and $\ket{p}$ decays at a rate $\gamma_{\downarrow p}$,  coming from the decay of state $\ket{p}$ to the ground state (by emitting photons
or by non-radiative channels via inelastic collisions) or from collisional dephasing; and the alkali-spin and noble-gas-spin coherences relax
at rates $\gamma_{\downarrow\uparrow}$ and $\gamma_{\Downarrow\Uparrow}$,
respectively, by various spin thermalization channels. In addition,
the atoms are moving, and their thermal motion is rendered diffusive
by the dense noble gas acting as a buffer \citep{Firstenberg-RMP}.
We denote by $D_{\mathrm{a}}$ and $D_{\mathrm{b}}$ the spatial diffusion
coefficients of the alkali and noble-gas atoms, respectively.

The overall dynamics can be described using the Heisenberg-Bloch-Langevin
formalism of open quantum systems \citep{Polzik-RMP-2010,Gorshkov1,CollectiveSpinStatesThermalDynamics}.
The atomic dynamics, in terms of the continuous quantum operators
$\hat{\sigma}_{\mu\nu}\left(\mathbf{r},t\right)$ in the Heisenberg
picture, is given by the stochastic differential equations
\begin{equation}
\partial_{t}\hat{\sigma}_{\mu\nu}=\frac{i}{\hbar}[\mathcal{H},\hat{\sigma}_{\mu\nu}]+(D_{\mathrm{a/b}}\nabla^{2}-\gamma_{\mu\nu})\hat{\sigma}_{\mu\nu}+\hat{f}_{\mu\nu}.\label{eq:Heisenberg_Langevin_general_form}
\end{equation}
The first term describes coherent evolution by the system Hamiltonian
$\mathcal{H}$. The second term describes the decay of the system
both due to the spatial diffusion and due to its coupling to the environment
at a rate $\gamma_{\mu\nu}$. The third term describes the stochastic
evolution through the input noise operators $\hat{f}_{\mu\nu}$, which
depend on the thermal spin-state of the reservoir \citep{Zoller-Quantum-Noise-1999}.
The explicit form of these equations is given in Eqs.~(\ref{eq:Bloch_no_approx_1})-(\ref{eq:Bloch_no_approx_3}).

\subsection{Collective excitations of polarized ensembles}

\begin{figure}[t]
\begin{centering}
\includegraphics[clip,width=7.5cm]{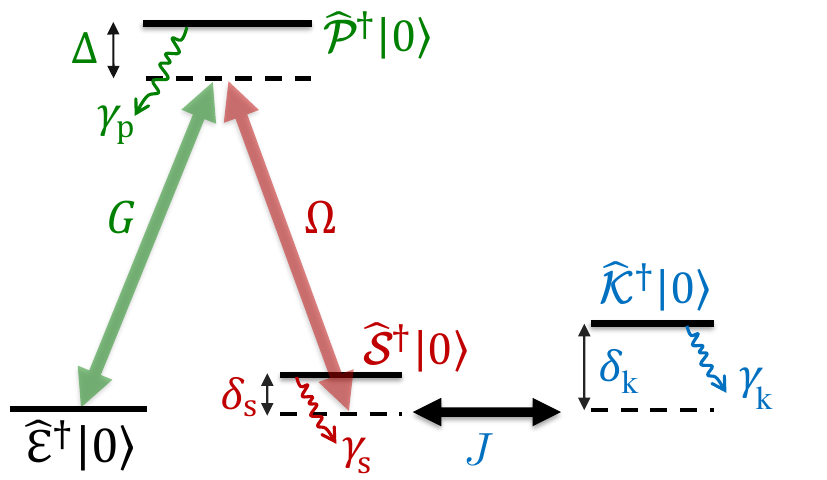}
\par\end{centering}
\centering{}\caption{Couplings between the collective states for a single input photon.
A single photon $\hat{\mathcal{E}}^{\dagger}|0\rangle$ in the cavity
is transferred to the collective noble-gas spin excitation $\hat{\mathcal{K}}^{\dagger}|0\rangle$,
which acts as the quantum memory, via intermediate excitations of
the alkali collective excited-state $\hat{\mathcal{P}}^{\dagger}|0\rangle$
and ground-state $\hat{\mathcal{S}}^{\dagger}|0\rangle$ spins. The
coupling is controlled by varying the strength of the control field
$\Omega(t)$ and the detunings $\delta_{\textnormal{s}}(t)$ and $\delta_{\textnormal{k}}(t)$.
The exchange coupling rate $J$ and the decay rates $\gamma_{\textnormal{p}},\,\gamma_{\textnormal{s}}$,
and $\gamma_{\textnormal{k}}$ are constant. The state $|0\rangle$
corresponds to the maximally polarized spin state, and $\hat{\mathcal{P}},\hat{\mathcal{S}}$,
and $\hat{\mathcal{K}}$ are bosonic field operators. \label{fig:singlephoton}}
\end{figure}

At this point, we focus on the regime of highly polarized spin ensembles,
where Eqs.~(\ref{eq:Hamiltonian_tot}), (\ref{eq:Hamiltonian_tot-1}),
and (\ref{eq:Heisenberg_Langevin_general_form}) can be further simplified
by using the Holstein-Primakoff transformation \citep{Polzik-RMP-2010,Holstein-Primakof}.
This transformation replaces the collective spin ladder operators
with bosonic creation and annihilation operators. Let $p_{\text{a}}$
and $p_{\text{b}}$ denote the polarization degree of the alkali and
noble-gas spin ensembles, respectively. For polarized spin ensembles
($p_{\text{a}},p_{\text{b}}\approx1$), the operators $\sigma_{\downarrow\downarrow}\approx p_{\text{a}}n_{\text{a}}$
and $\hat{\sigma}_{\Downarrow\Downarrow}\approx p_{\text{b}}n_{\text{b}}$
act as classical magnetic moments, where $n_{\text{a}}$ and $n_{\text{b}}$
are the alkali and noble-gas densities. The collective spin excitations,
which remain quantum, can now be described by the operators $\hat{\mathcal{P}}(\mathbf{r},t)=\hat{\sigma}_{\downarrow\textnormal{p}}(\mathbf{r},t)/\sqrt{p_{\text{a}}n_{\text{a}}}$,
$\hat{\mathcal{S}}(\mathbf{r},t)=\hat{\sigma}_{\downarrow\uparrow}(\mathbf{r},t)/\sqrt{p_{\text{a}}n_{\text{a}}}$, 
and $\hat{\mathcal{K}}(\mathbf{r},t)=\hat{\sigma}_{\Downarrow\Uparrow}(\mathbf{r},t)/\sqrt{p_{\text{b}}n_{\text{b}}}.$
These operators satisfy the bosonic commutation relations $[\hat{\mathcal{P}}(\mathbf{r}),\hat{\mathcal{P}}^{\dagger}(\mathbf{r}')]=[\hat{\mathcal{S}}(\mathbf{r}),\hat{\mathcal{S}}^{\dagger}(\mathbf{r}')]=[\hat{\mathcal{K}}(\mathbf{r}),\hat{\mathcal{K}}^{\dagger}(\mathbf{r}')]=\delta(\mathbf{r}-\mathbf{r}')$.
Figure \ref{fig:singlephoton} reformulates the level structure of
the system using these operators, for the case of a single excitation.

The equations of motion for the spin annihilation operators {[}see
Eqs.~(\ref{eq:Bloch_equation_ge})-(\ref{eq:Bloch_equation_mk}){]}
are then given by
\begin{align}
\partial_{t}\hat{\mathcal{P}}(\mathbf{r},t) & =-(\gamma_{\textnormal{p}}+i\Delta)\hat{\mathcal{P}}(\mathbf{r},t)+i\Omega(\mathbf{r},t)\hat{\mathcal{S}}(\mathbf{r},t)\label{eq:P_r_t_equation}\\
+ & iG(\mathbf{r})\hat{\mathcal{E}}\left(t\right)+\hat{f}_{\mathcal{P}}(\mathbf{r},t),\nonumber \\
\partial_{t}\hat{\mathcal{S}}(\mathbf{r},t) & =-(\gamma_{\textnormal{s}}+i\delta_{\text{s}}-D_{\textnormal{a}}\nabla^{2})\hat{\mathcal{S}}(\mathbf{r},t)\label{eq:S_r_t_equation}\\
+ & i\Omega^{*}(\mathbf{r},t)\hat{\mathcal{P}}(\mathbf{r},t)-iJ\hat{\mathcal{K}}(\mathbf{r},t)+\hat{f}_{\mathcal{S}}(\mathbf{r},t),\nonumber \\
\partial_{t}\hat{\mathcal{K}}(\mathbf{r},t) & =-(\gamma_{\textnormal{k}}+i\delta_{\text{k}}-D_{\text{b}}\nabla^{2})\hat{\mathcal{K}}(\mathbf{r},t)\label{eq:K_r_t_equation}\\
- & iJ\hat{\mathcal{S}}(\mathbf{r},t)+\hat{f}_{\mathcal{K}}(\mathbf{r},t).\nonumber
\end{align}
Here $\delta_{\text{s}}$ and $\delta_{\text{k}}$ are the two-photon detunings
associated with the collective spin operators of the alkali and noble
gas, respectively. They correspond to the previously-defined $\tilde{\delta}_{\textnormal{s}}$
and $\tilde{\delta}_{\textnormal{k}}$, but also include the collisional
shifts due to the spin-exchange collisions (see Appendix \ref{sec:Heisenberg-Bloch-Langevin-equati}). $G(\mathbf{r})=\sqrt{p_{\text{a}}n_{\text{a}}}g(\mathbf{r})$ denotes the collective interaction rate of the optical dipole with the optical field in the cavity, and
 $J=\zeta\sqrt{[I]p_{\text{\text{a}}}p_{\text{b}}n_{\text{a}}n_{\text{b}}/4}$
denotes the collective exchange rate of the two polarized spin ensembles,
a consequence of multiple weak spin-exchange collisions \citep{Firstenberg-Weak-collisions}.
The stochastic properties of the quantum noise
terms $\hat{f}_{\mathcal{P}}(\mathbf{r},t)$, $\hat{f}_{\mathcal{S}}(\mathbf{r},t)$
and $\hat{f}_{\mathcal{K}}(\mathbf{r},t)$ are detailed in Appendix
\ref{sec:Properties-of-Quantum}.

Equations (\ref{eq:P_r_t_equation})-(\ref{eq:K_r_t_equation}) manifest
the collective enhancement of the various couplings to the collective
excitations. The atom-photon interaction rate $g(\mathbf{r})$ is multiplied
by the large factor $\sqrt{n_{\text{a}}}$, as expected for the coherent
absorption and emission of a photon by multiple atoms, resulting with the enhanced collective rate $G(\mathbf{r})$. The microscopic
coherent spin-exchange rate $\zeta$ is multiplied by $\sqrt{n_{\text{a}}n_{\text{b}}}$,
corresponding to a unitary precession of the collective spin of one
gas around the other at the enhanced rate $J$.

\subsection{Dynamics of the light field}

The dynamics of the slowly-varying quantum light field in the cavity
$\hat{\mathcal{E}}\left(t\right)$ is described by the Heisenberg-Langevin
equation
\begin{align}
\partial_{t}\hat{\mathcal{E}} & =-\kappa\mathcal{\hat{E}}+\sqrt{2\kappa}\hat{\mathcal{E}}_{\text{in}}+\frac{i}{\hbar}[\mathcal{H},\hat{\mathcal{E}}]\label{eq:cavity_field_equation}\\
 & =-\kappa\hat{\mathcal{E}}+\sqrt{2\kappa}\hat{\mathcal{E}}_{\text{in}}+i\int_{V}G^{*}\left(\mathbf{r}\right)\hat{\mathcal{P}}(\mathbf{r},t)d^{3}\mathbf{r}.\nonumber
\end{align}
Here the cavity field decays at a rate $\kappa$ and is driven by the field $\hat{\mathcal{E}}_{\text{in}}\left(t\right)$
at the cavity input port.
The third term in Eq.~(\ref{eq:cavity_field_equation}) describes
the collective absorption and emission of the field by the dipole
coherence $\hat{\mathcal{P}}$. We implicitly assume that the cavity
has no internal losses. The field $\hat{\mathcal{E}}_{\text{out}}\left(t\right)$
at the cavity output port is obtained from  the general input-output
relation \citep{Gorshkov1}
\begin{equation}
\hat{\mathcal{E}}_{\text{out}}=\sqrt{2\kappa}\hat{\mathcal{E}}-\hat{\mathcal{E}}_{\text{in}}.\label{eq:Input_Output_relation}
\end{equation}
The commutation relations $[\hat{\mathcal{E}}_{\text{in}}\left(t\right),\hat{\mathcal{E}}_{\text{in}}^{\dagger}\left(t'\right)]=[\hat{\mathcal{E}}_{\text{out}}\left(t\right),\hat{\mathcal{E}}_{\text{out}}^{\dagger}\left(t'\right)]=\delta\left(t-t'\right)$
hold. In the fast-cavity regime ($\kappa\gg G$,
also known as a 'bad' cavity), the input-output relation simplifies
to
\begin{equation}
\hat{\mathcal{E}}_{\text{out}}=\hat{\mathcal{E}}_{\text{in}}+i\sqrt{\frac{2}{\kappa}}\int_{V}G^{*}\left(\mathbf{r}\right)\hat{\mathcal{P}}\left(\mathbf{r},t\right)d^{3}\mathbf{r}.\label{eq:Output_field_cavity}
\end{equation}
We adopt this approximation in our analysis, thus limiting the results
to the fast-cavity regime. Notably, the operation of an optical quantum
memory in this regime resembles the operation in free space \citep{Gorshkov2}.

To further simplify the analysis, we limit the bandwidth of the incoming
light pulse to
\begin{equation}
T^{-1}\ll C\gamma_{\textnormal{p}},\label{eq:adiabatic_condition_P}
\end{equation}
where $T$ is the pulse duration. The cooperativity parameter of the
cavity is given by
\begin{equation}
C=\frac{\left|G\right|^{2}}{\gamma_{\textnormal{p}}\kappa},
\end{equation}
characterizing the collective atom-photon interaction strength with
respect to the decay rates, where $G=\sqrt{p_{\text{a}}n_{\text{a}}}g$  and $g$ is given in Eq.~(\ref{eq:definition of G}).
Under the assumptions of fast cavity and limited pulse bandwidth,
$\hat{\mathcal{P}}$ adiabatically follows both the light field $\hat{\mathcal{E}}$
and the collective alkali-spin operators $\hat{\mathcal{S}}$, satisfying
\begin{equation}
\hat{\mathcal{P}}(\mathbf{r},t)=i\frac{\Omega(\mathbf{r},t)\hat{\mathcal{S}}(\mathbf{r},t)+G(\mathbf{r})\hat{\mathcal{E}}\left(t\right)-i\hat{f}_{\mathcal{P}}(\mathbf{r},t)}{\gamma_{\textnormal{p}}+i\Delta}.\label{eq:P_local_adiabatic}
\end{equation}

\section{Spatial modes of atomic operators}\label{sec:spatial_modes}

Up until this point, we described the dynamics of the atomic spins
using local continuous operators. Indeed, Eqs.~(\ref{eq:S_r_t_equation}),
(\ref{eq:K_r_t_equation}), and (\ref{eq:P_local_adiabatic}) support
the storage of photons in multiple spatial modes \citep{Firstenberg-RMP}.
The signal light field, however, was assumed to reside in the specific
spatial mode $f_{\varepsilon}\left(\mathbf{r}\right)$ of the cavity,
with the input signal field matching this mode. Therefore, the signal
excites an atomic superposition with a particular spatial amplitude
pattern {[}the term $\propto G\left(\mathbf{r}\right)\hat{\mathcal{E}}\left(t\right)$
in Eq.~(\ref{eq:P_local_adiabatic}){]}, and subsequently this specific
superposition coherently emits light to the output port {[}Eq.~(\ref{eq:Output_field_cavity}){]}.
{The general multi-mode evolution of the collective spins $\hat{\mathcal{S}}\left(\boldsymbol{r},t\right)$ and $\hat{\mathcal{K}}\left(\boldsymbol{r},t\right)$ during  memory operation is governed by the spatial profiles of the signal and control fields, as well as by the nonlocal action of the diffusion operator. Specifically, $\hat{\mathcal{S}}\left(\boldsymbol{r},t\right)$ is driven via a two-photon transition by the control and signal fields, generating a quantum excitation of the collective alkali spin-wave with a long wavelength. In this section, we present simplified equations of motion which consider the excitation of only the single uniform modes $\hat{\mathcal{S}}\left(t\right)$ and $\hat{\mathcal{K}}\left(t\right)$ of $\hat{\mathcal{S}}\left(\boldsymbol{r},t\right)$ and $\hat{\mathcal{K}}\left(\boldsymbol{r},t\right)$, whereas, in Appendix \ref{sec:Spatial-modes-representation},  we derive the general multi-mode equations of motion and, in Appendix \ref{sec:Diffusion-relaxation-rate}, present numerically calculated solutions.}

In the single-mode representation, the output field of the cavity
is given by {[}compare to Eq.~(\ref{eq:Output_field_cavity}){]}
\begin{equation}
\hat{\mathcal{E}}_{\text{out}}\left(t\right)=\hat{\mathcal{E}}_{\text{in}}\left(t\right)+i\sqrt{2C\gamma_{\textnormal{p}}}\hat{\mathcal{P}}\left(t\right),\label{eq:output-light-P}
\end{equation}
where the atomic optical dipole is {[}compare to Eq.~(\ref{eq:P_local_adiabatic}){]}
\begin{equation}
\hat{\mathcal{P}}\left(t\right)=\frac{i\Omega(t)\hat{\mathcal{S}}\left(t\right)+i\sqrt{2\gamma_{\textnormal{p}}C}\hat{\mathcal{E}}_{\text{in}}\left(t\right)+\hat{f}_{\mathcal{P}}\left(t\right)}{\gamma_{\textnormal{p}}\left(1+C\right)+i\Delta}.\label{eq:P adiabatic-1}
\end{equation}
The dynamics of the uniform mode $\hat{\mathcal{S}}\left(t\right)$ of the alkali
spin is given by
\begin{align}
\partial_{t}\hat{\mathcal{S}}= & -(\gamma_{\text{s}}+\Gamma_{\Omega}+i\delta_{\text{s}})\hat{\mathcal{S}}-iJ\hat{\mathcal{K}}-Q\Omega^{*}\mathcal{\hat{E}}_{\text{in}}+\hat{F}_{\mathcal{S}}\label{eq:Uniform_alkali_spin_mode_diff_equation}
\end{align}
where the complex-valued optical coupling rate is
\begin{equation}
\Gamma_{\Omega}\left(t\right)\equiv\frac{|\Omega(t)|^{2}}{\gamma_{\textnormal{p}}\left(1+C\right)+i\Delta},\label{eq:R- power broadening rate}
\end{equation}and we define 
\begin{equation}
Q=\frac{\sqrt{2C\gamma_{\textnormal{p}}}}{\gamma_{\textnormal{p}}\left(1+C\right)+i\Delta}.\label{eq:Q_definition}
\end{equation}We identify $\gamma_{\Omega}\equiv\text{re}(\Gamma_{\Omega})$ as the stimulated (power-broadened) optical coupling rate to the alkali
spins, and $\text{im}(\Gamma_{\Omega})$ as the light shift due to the control field. The noise operator of the alkali spins is given by $\hat{F}_{\mathcal{S}}=\hat{f}_{\mathcal{S}}+iQ\Omega^{*}\hat{f}_{\mathcal{P}}/\sqrt{2C\gamma_{\textnormal{p}}}$, including the excess noise due to scattering of the control photons. 

The uniform mode of the alkali spins has a large overlap with the long-lived uniform mode of the noble-gas spins, which is unaffected by diffusion \citep{CollectiveSpinStatesThermalDynamics} and therefore chosen as the quantum memory.
The dynamics of the uniform mode $\hat{\mathcal{K}}\left(t\right)$
of the noble-gas spin is given by
\begin{equation}
\partial_{t}\hat{\mathcal{K}}=-(\gamma_{\textnormal{k}}+i\delta_{\text{k}})\hat{\mathcal{K}}-iJ\hat{\mathcal{S}}+\hat{f}_{\mathcal{K}}.\label{eq:Uniform_noble_spin_mode_diff_equation}
\end{equation}
The noise operators $\hat{f}_{\mathcal{P}},\hat{f}_{\mathcal{S}}$,
and $\hat{f}_{\mathcal{K}}$ are defined in Appendix \ref{sec:Spatial-modes-representation}.

\section{Memory efficiency}\label{sec:mem_eff}

Following Refs.~\citep{Gorshkov1,Gorshkov-PRL}, we write the total
efficiency of the quantum memory
\begin{equation}
\eta_{\text{tot}}=\eta_{\text{in}}\eta_{\text{dark}}\eta_{\text{out}}\label{eq:Total efficiency}
\end{equation}
in terms of the efficiency $\eta_{\text{in}}$ of the storage process, the noble-gas efficiency in the dark $\eta_{\text{dark}}$
the efficiency $\eta_{\text{out}}$ of the retrieval process. The total efficiency $\eta_{\text{tot}}$
sets a limit on other figures of merit, such as the memory fidelity
\citep{Gorshkov1} or preservation of squeezing \citep{Dantan1,Dantan2,Dantan3}.

{We assume that the signal pulse extends from $t=-\infty$ to $t=0$. The storage process we consider occurs during the time interval $-\infty \leq t \leq T'$, where the duration $T'\geq 0$  enables further manipulation of the spins in the absence of the signal field to complete the mapping.} In this process, the input field is mapped onto
the long-lived, collective, noble-gas spin $\hat{\mathcal{E}}_{\text{in}}\rightarrow\hat{\mathcal{K}}(T')$.
We denote $\langle\hat{\mathcal{O}}^\dagger\hat{\mathcal{O}}\rangle_{t}=\langle\hat{\mathcal{O}}^\dagger(t)\hat{\mathcal{O}}(t)\rangle$ for any operator $\hat{\mathcal{O}}$ and define the storage efficiency by the ratio
\begin{equation}
\eta_{\text{in}}\equiv\frac{\langle\hat{\mathcal{K}}^\dagger\hat{\mathcal{K}}\rangle_{T'}}{\int_{-\infty}^{0}\langle\hat{\mathcal{E}}^\dagger_{\text{in}}\hat{\mathcal{E}}_{\text{in}}\rangle_{t} dt},\label{eq:Storage efficiency definition}
\end{equation}
\emph{i.e.,} by the number of stored spin excitations divided by the number of incoming signal photons.
We can use the integral relation in Eq.~(\ref{eq:Integral excitation conservation})
to express the efficiency as
\begin{align}
\eta_{\text{in}}= & 1-\frac{2\int_{-\infty}^{T'}dt\bigl\langle\tfrac{1}{2}\hat{\mathcal{E}}_{\text{out}}^\dagger\hat{\mathcal{E}}_{\text{out}}+\gamma_{\textnormal{p}}\hat{\mathcal{P}}^\dagger\hat{\mathcal{P}}+\gamma_{\text{s}}\hat{\mathcal{S}}^\dagger\hat{\mathcal{S}}+\gamma_{\text{k}}\hat{\mathcal{K}}^\dagger\hat{\mathcal{K}}\bigl\rangle_t}{\int_{-\infty}^{0}dt\langle\hat{\mathcal{E}}_{\text{in}}^\dagger\hat{\mathcal{E}}_{\text{in}}\rangle_{t}}.\label{eq:storage efficiency}\end{align}
We find that the storage efficiency is limited by four relaxation
mechanisms: decoherence of excited alkali atoms at a rate $\gamma_{\textnormal{p}}$,
decoherence of the alkali-metal spin at a rate $\gamma_{\text{s}}$,
decoherence of the noble-gas spin at a rate $\gamma_{\text{k}}$,
and leakage of photons during the storage represented by $\int_{-\infty}^{T'}\langle\hat{\mathcal{E}}^\dagger_{\text{out}}\hat{\mathcal{E}}_{\text{out}}\rangle_{t} dt$.

After the storage stage, the control fields are fixed and the memory preserves the noble-gas spin coherence for a memory (dark) time $\tau$. The efficiency 
of this stage is determined by the noble-gas relaxation in the dark and is given by
\begin{equation}
\eta_{\text{dark}}=\frac{\langle\hat{\mathcal{K}}^\dagger\hat{\mathcal{K}}\rangle_{(T'+\tau)}}{\langle\hat{\mathcal{K}}^\dagger\hat{\mathcal{K}}\rangle_{T'}}=\exp(-2\gamma_{\text{k}}\tau).\label{eq:relaxation in the dark noble gas}
\end{equation}

The retrieval process starts at $t=T'+\tau$. During this process, there is no input field, and the spin excitations
are mapped to the output field $\hat{\mathcal{K}}(T'+\tau)\rightarrow\hat{\mathcal{E}}_{\textnormal{out}}$. 
The retrieval ends when no atomic excitations are left in the medium.
The retrieval efficiency is then given by
\begin{equation}
\eta_{\text{out}}\equiv\frac{\int_{\tau+2T'}^{\infty}\langle\hat{\mathcal{E}}^\dagger_{\text{out}}\hat{\mathcal{E}}_{\text{out}}\rangle_{t} dt}{\langle \hat{\mathcal{K}}^\dagger \hat{\mathcal{K}} \rangle_{(T'+\tau)}},\label{eq:retrieval efficiency definition}
\end{equation}
\emph{i.e.}, the number of retrieved photons divided by the number
of stored spin excitations. Similar to the storage process, we assume that the control field starts only $2T'+\tau$, delayed by a duration $T'\geq0$ with respect to the retrieval start. This delay enables manipulation of the spin-gases in the dark before the control field is able to initiates the emission of the optical signal. Using the integral relation in Eq.~(\ref{eq:Integral excitation conservation}),
we find
\begin{equation}
\eta_{\text{out}}=1-\frac{2\int_{T'+\tau}^{\infty}\bigl\langle\gamma_{\textnormal{p}}\hat{\mathcal{P}}^\dagger\hat{\mathcal{P}}+\gamma_{\text{s}}\hat{\mathcal{S}}^\dagger \hat{\mathcal{S}}+\gamma_{\text{k}}\hat{\mathcal{K}}^\dagger \hat{\mathcal{K}}\bigl\rangle_t dt}{\langle\hat{\mathcal{K}}^\dagger\hat{\mathcal{K}}\rangle_{(T'+\tau)}}.\label{eq:retreival efficiency}
\end{equation}
 It is evident that the retrieval efficiency is maximized if the duration
for which $\hat{\mathcal{P}}$ and $\hat{\mathcal{S}}$ are excited
is minimal. 

\section{Protocols for light storage\label{sec:Numerical-storage}}

Our motivation for setting up an interface between optical signals and
noble-gas spin ensembles is to utilize the long-lived uniform mode of collective states of the noble-gas to store the quantum state of an input optical field. In other words, we aim to find controllable processes which 
transfer quantum excitations from $\hat{\mathcal{E}}_{\text{in}}(t)$
to $\hat{\mathcal{K}}(T')$ efficiently and then enable controllable retrieval of the light field by mapping  $\hat{\mathcal{K}}(T'+\tau)$ to $\hat{\mathcal{E}}_{\text{out}}(t)$.

Here we present and exemplify the memory operation using two distinctive protocols, which attain high efficiencies in complementary regimes. The first protocol, presented in Sec.~\ref{sec:Fast-storage-on} temporarily maps the optical signal onto the three-level alkali memory, which accepts higher bandwidth, and only later transfers it to the noble-gas spins via the spin-exchange coupling. We term this two-step approach as sequential mapping and demonstrate that it is an efficient protocol when the optical signal pulse is shorter than the the alkali decoherence time, and when the spin-exchange coupling between the two gases is strong. The second protocol, presented in Sec.~\ref{sec:Adiabatic-storage-on}, constitutes a direct mapping between the light and noble-gases, using the alkali spins as mediators that adiabatically follow, keeping their excitation small. This protocol is shown efficient when the optical pulse is longer than the typical exchange time between the gases. In Sec.~\ref{sec:comparison_seq_adi} we compare the provide a comparison of these two protocols. 

{To simplify the presentation, we exemplify the protocols using an input signal that has an exponential temporal profile of the form
\begin{equation}
\mathcal{E}_{{\rm in}}(t)=\sqrt{\frac{2}{T}}e^{t/T},\,\,\,\,\,\,\,\:\,\,-\infty\leq t\leq 0,\label{eq:exponentially shaped pulse input}
\end{equation}
and zero otherwise. We also assume that the desired retrieved field follows an exponential temporal profile of the form \begin{equation}
\mathcal{E}_{{\rm out}}(t)=\sqrt{\frac{2}{T}}e^{(2T'+\tau-t)/T},\,\,\,\,\,\,\,\:\,\, t\geq 2T'+\tau.\label{eq:exponentially shaped pulse output}
\end{equation}
 Here $T$ denotes the characteristic pulse duration of the input and retrieved light, which determines the optical bandwidth of the pulse $1/(2T)$. This particular choice of pulse profiles enables considerable analytic simplification of the theoretical results. Indeed, for $\Lambda$-type memories in the adiabatic regime, optimal storage and retrieval of such exponentially-shaped signals is associated with temporally square-shaped control pulses \citep{Gorshkov1}. Therefore, this choice allows us to derive simple expressions for the efficiencies, with the control field $\Omega(t)$ [or equivalently $\gamma_{\Omega}(t)$] kept constant for the duration of the signal pulse. Solutions for other pulse profiles might require the application of optimal control tools as those presented in Ref.~\citep{Gorshkov4}. Nevertheless, the particular solution presented here enables simple identification of the main competing mechanisms in the operation of the memory  and, in particular, highlights the effect of the characteristic pulse duration $T$ on memory efficiency.}

\subsection{Sequential mapping\label{sec:Fast-storage-on}}
We first analyze the sequential protocol, which is suitable for storage and retrieval of short pulses in which $T\ll1/\gamma_{\textnormal{s}}$ and $T\lesssim1/J$.  
In this protocol, storage is conducted by a sequential two-stage transfer of the excitation
$\hat{\mathcal{E}}_{\text{in}}\rightarrow\hat{\mathcal{S}}(0)\rightarrow\hat{\mathcal{K}}(T')$,
by first mapping the signal onto the alkali spin $\hat{\mathcal{S}}$
at time $t=0$, and afterwards mapping it onto the noble-gas spin $\hat{\mathcal{K}}$
at time $T'$. After the memory time of duration $\tau$ in which the noble-gas stores the signal's excitations, they are retrieved into emission of an optical signal by following the two stage sequence $\hat{\mathcal{K}}(T'+\tau)\rightarrow\hat{\mathcal{S}}(2T'+\tau)\rightarrow\hat{\mathcal{E}}_{\text{out}}$.

\begin{figure*}[t]
\begin{centering}
\includegraphics[scale=0.55]{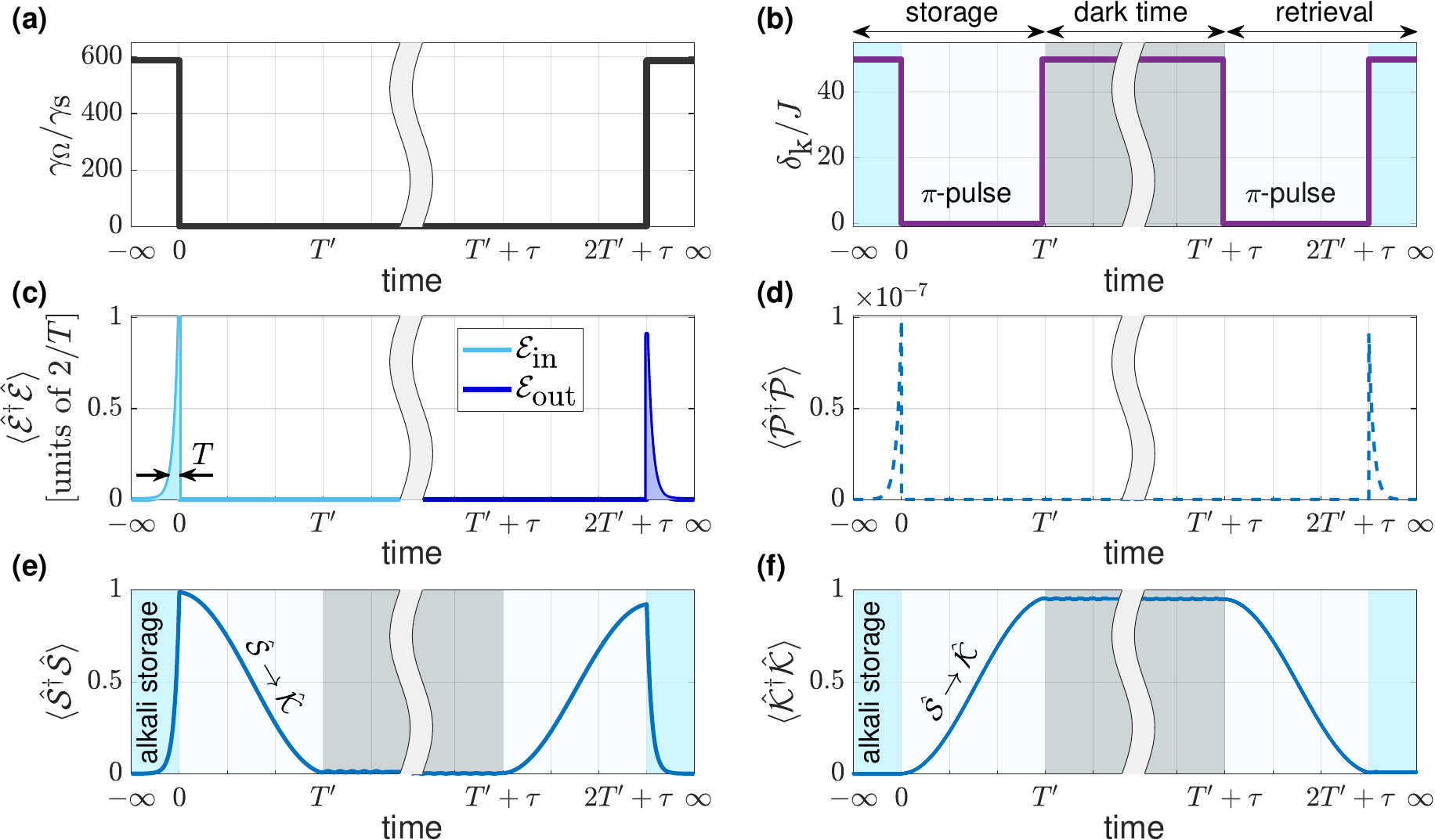}
\par\end{centering}
\centering{}\caption{{Example of storage and retrieval of a short signal pulse using the sequential mapping protocol. We solve numerically Eqs.\ (\ref{eq:Uniform_alkali_spin_mode_diff_equation}) and (\ref{eq:Uniform_noble_spin_mode_diff_equation}) for an exponentially-shaped input signal field with a single excitation [cf.~Eq.~(\ref{eq:exponentially shaped pulse input})], setting $\Delta=\delta_{\textrm{s}}=0$ and using square pulses for the optical control field $\gamma_\Omega$ and magnetic detuning $\delta_{\mathrm{k}}$ as shown in (a-b). Here $J=60\gamma_{\textnormal{s}}$,  $\gamma_{\textnormal{s}}T=1.7\cdot 10^{-3}$, and $C=100$. The memory maps the excitation of the input field $\langle\hat{\mathcal{E}}_{\textnormal{in}}^\dagger\hat{\mathcal{E}}_{\textnormal{in}}\rangle_{t}$ onto the atomic populations of the alkali $\langle\hat{\mathcal{P}}^\dagger\hat{\mathcal{P}}\rangle_{t}$, $\langle\hat{\mathcal{S}}^\dagger\hat{\mathcal{S}}\rangle_{t}$,
and the noble gas $\langle\hat{\mathcal{K}}^\dagger\hat{\mathcal{K}}\rangle_{t}$, which are later retrieved, on-demand, into an optical excitation of $\langle\hat{\mathcal{E}}_{\textnormal{out}}^\dagger\hat{\mathcal{E}}_{\textnormal{out}}\rangle_{t}$. In the first stage
$-\infty\le t\le 0$, the control field maps the signal onto the collective state of the alkali spins via a two-photon process, while
the noble gas remains decoupled by setting a large $\delta_{\textnormal{k}}$.
In the second stage {$0<t\le T'$} (here $T'/T=15.6$), by tuning $\delta_{\textnormal{k}}=0$, the alkali spins exchange the excitation with the noble-gas spins.
Subsequently, the noble-gas spins are decoupled from the alkali spins
by increasing $\delta_{\textnormal{k}}$ again. The optical signal is retrieved into the exponentially shaped output mode [cf.~Eq.(\ref{eq:exponentially shaped pulse output})]
by time reversing the storage sequence, except for a correction of order $\gamma_{\textnormal{s}}T$ to the control field amplitude, $\gamma_{\Omega }(\textnormal{retrieval})=\gamma_{\Omega}(\textnormal{storage})-2\gamma_{\text{s}}$,  which corrects for the effect of nonzero alkali-spin relaxation to first order, as described in the text. The calculated memory efficiency is $\eta_{\textrm{tot}}=0.91$.} \label{fig:fast_Storage}}
\end{figure*}

{To exemplify this protocol, we first solve numerically Eqs.~(\ref{eq:Uniform_alkali_spin_mode_diff_equation}) and (\ref{eq:Uniform_noble_spin_mode_diff_equation}) as ordinary differential equations by setting the noise operators to zero. We use the input field in Eq.~(\ref{eq:exponentially shaped pulse input}) and solve the storage and retrieval stages for constant $\Delta=\delta_{\rm s}=0$ and for $\gamma_{\Omega}(t),\delta_{\rm k}(t)$ that follow square temporal profiles as shown in Fig.~\ref{fig:fast_Storage}(a-b). In Fig.~\ref{fig:fast_Storage}(c-f), we present the calculated quantum excitations of the optical input and output fields and the atomic excitations $|\mathcal{P}|^{2}$, $|\mathcal{S}|^{2}$,
and $|\mathcal{K}|^{2}$ using
 $J=60\gamma_{\textnormal{s}},\,\gamma_{\textnormal{s}}T=1.7\cdot 10^{-3}$, and $C=100$, setting initially the atomic excitations to zero. The retrieved optical field is presented in Fig.~\ref{fig:fast_Storage}(c). The overall calculated memory efficiency in this example is  $\eta_{\textrm{tot}}=0.91$.}
 
 Below we examine the protocol's performance analytically and consider its efficiency as a function of $J,\gamma_{\rm s}$, and $T$. We simplify the discussion by neglecting the slow relaxation of the noble-gas spins (\emph{i.e.}, setting $\gamma_\text{k}=0$) during the mapping processes and retain $\gamma_\text{k}$ only for the long memory time in the dark.

\subsubsection{Storage stage 1: $\hat{\mathcal{E}}_\text{in}\rightarrow\hat{\mathcal{\boldsymbol{S}}}$}

In the first stage the signal field is mapped onto the alkali spins. To achieve this $\delta_{\text{k}}$ is kept large, which efficiently decouples the noble-gas spins from the alkali spins, thus rendering the first storage
stage similar to standard light storage onto alkali spins. Indeed, for $\delta_{\text{k}}\gg J,T^{-1}$
we find from Eq.~(\ref{eq:Uniform_noble_spin_mode_diff_equation})
that $\hat{\mathcal{K}}\approx(J/\delta_{\text{k}})\hat{\mathcal{S}}$,
such that $\langle \hat{\mathcal{K}}^\dagger \hat{\mathcal{K}}
\rangle\ll1$.

In Appendix \ref{sec:alkali_memory}, we review the process of light
storage onto and retrieval from alkali spins, following Ref.~\citep{Gorshkov1}.
The maximal efficiency of the process  $\hat{\mathcal{E}}_{\text{in}}\rightarrow\hat{\mathcal{S}}$ is given in Eq.~(\ref{eq:3-level-efficiency}) for a general pulse shape and optimally shaped control. For the particular case of exponentially-shaped signal we consider here, a square control pulse  with a constant value of $\gamma_{\Omega}=1/T+\gamma_{\text{s}}$ maximizes the storage efficiency, which reads

\begin{equation}
\eta_{\text{in}}^{(\mathcal{E}\rightarrow\mathcal{S})}=\frac{\langle\hat{\mathcal{S}}^\dagger\hat{\mathcal{S}}\rangle_{\left(t=0\right)}}{\int_{-\infty}^{0}\langle\hat{\mathcal{E}}^\dagger_{\text{in}}\hat{\mathcal{E}}_{\text{in}}\rangle_{t} dt}=\frac{C}{C+1}\cdot\frac{1}{1+\gamma_{\text{s}}T}.\label{eq:Storage-efficiency-alkali}
\end{equation}
\subsubsection{Storage stage 2: $\hat{\mathcal{\boldsymbol{S}}}\rightarrow\hat{\mathcal{\boldsymbol{K}}}$}

In the second stage, as shown in Fig.~\ref{fig:fast_Storage}, the
control field is turned off ($\Omega=0$), and the excitation is coherently
transferred from the alkali to the noble-gas spins $\hat{\mathcal{S}}(0)\rightarrow\hat{\mathcal{K}}(T')$
via the spin-exchange interaction. {In general, the dynamics of spin-exchange
in the presence of diffusion is multi-mode, due to thermal motion of the atoms during the exchange process \citep{Firstenberg-Weak-collisions,CollectiveSpinStatesThermalDynamics},
and we present the complete multi-mode calculation in Appendix \ref{sec:multi-mode-sequential}.}
Here we present the approximated dynamics for the uniform modes. This
solution is accurate for anti-relaxation coated cells \citep{CollectiveSpinStatesThermalDynamics}.

The exchange evolution of the alkali and noble-gas spins over the
duration $T'$ in the absence of light fields is given by \citep{Firstenberg-Weak-collisions}
\begin{align}
\hat{\mathcal{S}}(T') & =e^{-\frac{1}{2}(i\delta_{\text{s}}+i\delta_{\text{k}}+\gamma_{\text{s}}) T'}\biggl\{-\frac{iJ}{\tilde{J}}\sin(\tilde{J} T')\hat{\mathcal{K}}(0)\label{eq:alkali-noble-gas-uniform-exchange}\\
+\Bigl[ & \cos(\tilde{J} T')-\frac{\gamma_{\text{s}}-i\delta}{2\tilde{J}}\sin(\tilde{J} T')\Bigr]\hat{\mathcal{S}}(0)\biggr\}+\mathcal{\hat{W}}_{\text{s}}( T')\nonumber
\end{align}
for the alkali spins and
\begin{align}
\hat{\mathcal{K}}(T') & =e^{-\frac{1}{2}(i\delta_{\text{s}}+i\delta_{\text{k}}+\gamma_{\text{s}}) T'}\biggl\{-\frac{iJ}{\tilde{J}}\sin(\tilde{J} T')\hat{\mathcal{S}}(0)\label{eq:noble-gas-spin-evolution}\\
+\Bigl[ & \cos(\tilde{J} T')+\frac{\gamma_{\text{s}}-i\delta}{2\tilde{J}}\sin(\tilde{J} T')\Bigr]\hat{\mathcal{K}}(0)\biggr\}+\mathcal{\hat{W}}_{\text{k}}( T')\nonumber
\end{align}
for the noble-gas spins. Here
\begin{equation}
\delta=\delta_{\text{k}}-\delta_{\text{s}}\label{eq:relative_delta}
\end{equation}
is the mismatch between the spin precession frequencies, and \begin{equation}
\tilde{J}=\sqrt{J^{2}+(\delta+i\gamma_{\text{s}})^{2}/4}\label{eq:effective-exchange-rate}
\end{equation}
denotes the effective exchange rate. The stochastic quantum processes
$\mathcal{\hat{W}}_{\text{s}}$ and $\mathcal{\hat{W}}_{\text{k}}$
for the alkali and noble-gas spins are defined in Appendix \ref{sec:multi-mode-sequential}.

Maximal exchange rate and thus efficient mapping $\hat{\mathcal{S}}(0)\rightarrow\hat{\mathcal{K}}(T')$
are obtained by setting $\delta(B)=0$ during the exchange time $T'$. To simplify the analysis at this point, we consider the particular regime of strong coupling $J\gg\gamma_{\text{s}}$, which enables efficient exchange between the spin gases. In this regime, it is efficient to set $T'$ as the $\pi$-pulse duration, which is given by $T'\approx(\pi\tilde{J}-\gamma_{\textnormal{s}})/(2\tilde{J}^{2})$ 
to leading order in $\gamma_{\text{s}}/\tilde{J}$, yielding
\begin{equation}\label{eq:K_exchange}
\hat{\mathcal{K}}(T')=e^{-\frac{\pi\gamma_{\text{s}}}{4J}}\hat{\mathcal{S}}(0)+\mathcal{\hat{W}}_{\text{k}}(T').
\end{equation} Substitution of Eq.~(\ref{eq:K_exchange}) into Eq.~(\ref{eq:retrieval efficiency definition})
yields the storage efficiency for the second stage
\begin{equation}
\eta_{\text{in}}^{(\mathcal{S}\rightarrow\mathcal{K})}=\exp\left(-\frac{\pi\gamma_{\text{s}}}{2J}\right).\label{eq:sequential-memory-efficiency-stage-2}
\end{equation}

\subsubsection{Memory time}

Once the excitation is stored on the uniform mode of the noble-gas
spins, we can decouple the state of the two spin ensembles by applying
a large magnetic field. {Specifically, we take  $\delta\left(B\right)\gg J$, such that the generalized exchange rate   of Eq.~(\ref{eq:effective-exchange-rate}) becomes $\tilde{J}\approx\delta/2$. As the exchange contrast of collective spin excitations between the alkali ($\langle \hat{\mathcal{S}}^\dagger \hat{\mathcal{S}}\rangle$) and noble gas ($\langle \hat{\mathcal{K}}^\dagger \hat{\mathcal{K}}\rangle$) scales as $|J/\tilde{J}|^2$, for $\delta \gg J$ the exchange is negligible.} Under these conditions, Eq.~(\ref{eq:noble-gas-spin-evolution}) 
becomes
\begin{equation}
\hat{\mathcal{K}}(T'+\tau)=e^{-(i\delta_{\text{k}}+\gamma_{\text{k}})\tau}\hat{\mathcal{K}}(T')+\mathcal{\hat{W}}'
\end{equation}
for the memory time $\tau$, where $\mathcal{\hat{W}}'=\mathcal{\hat{W}}_{\text{k}}(T'+\tau)-\mathcal{\hat{W}}_{\text{k}}(T')$.
The noble-gas spins then act as a quantum memory that decays according
$\eta_{\mathrm{dark}}(\tau)$ in Eq.~(\ref{eq:relaxation in the dark noble gas}),
with potentially very long lifetime $\gamma_{\text{k}}^{-1}$.

\subsubsection{Retrieval: $\hat{\mathcal{\boldsymbol{K}}}\rightarrow\hat{\mathcal{\boldsymbol{S}}}\rightarrow\hat{\mathcal{E}}_{\text{out}}$}

We retrieve the photons from the memory by realizing the process $\hat{\mathcal{K}}(T'+\tau)\rightarrow\hat{\mathcal{S}}(2T'+\tau)\rightarrow\hat{\mathcal{E}}_{\text{out}}$.
First, the magnetic field is tuned to strongly couple the two spin
gases by setting $\delta\left(B\right)=0$. As described by Eq.~(\ref{eq:alkali-noble-gas-uniform-exchange}),
the excitation is mapped back from the noble gas to the alkali spins
after the same transfer time $T'$
\begin{equation}
\langle \hat{\mathcal{S}}^\dagger  \hat{\mathcal{S}}\rangle _{t_\text{r}}=e^{-\frac{\pi\gamma_{\text{s}}}{2J}}\langle \hat{\mathcal{K}}^\dagger \hat{\mathcal{K}}\rangle_{(T'+\tau)},
\end{equation} where $t_{\text{r}}=2T'+\tau$, yielding the retrieval efficiency of $\eta_{\text{out}}^{(\mathcal{K}\rightarrow\mathcal{S})}=\eta_{\text{in}}^{(\mathcal{S}\rightarrow\mathcal{K})}$.

Retrieval of the photons from the alkali spins is then performed as a standard retrieval in $\Lambda$-system memories, as reviewed in Appendix \ref{sec:alkali_memory}. Importantly, the temporal profile of the control field determines the emission profile of the retrieved signal field \citep{Gorshkov1}. 
For short pulses $T < 1/\gamma_s$, retrieval into the target mode we consider in Eq.~(\ref{eq:exponentially shaped pulse output}) is realized using a constant control field with $\gamma_{\Omega }=1/T-\gamma_{\text{s}}$ [compare to $\gamma_{\Omega}=1/T+\gamma_{\text{s}}$ during storage]. This profile yields  the retrieval  efficiency\begin{equation}
\eta_{\text{out}}^{(\mathcal{S}\rightarrow\mathcal{E})}=\frac{\int_{t_\text{r}}^{\infty}\langle\hat{\mathcal{E}}_{\text{out}}^\dagger \hat{\mathcal{E}}_{\text{out}}\rangle_t dt}{\langle\hat{\mathcal{S}}^\dagger\hat{\mathcal{S}}\rangle_{t_\text{r}}}=\frac{C}{C+1}(1-\gamma_{\text{s}}T)\label{eq:Ret-efficiency-alkali}
\end{equation}
[compare to Eq.~(\ref{eq:Storage-efficiency-alkali}) for storage].

\begin{figure*}[t]
\begin{centering}
\includegraphics[scale=0.55]{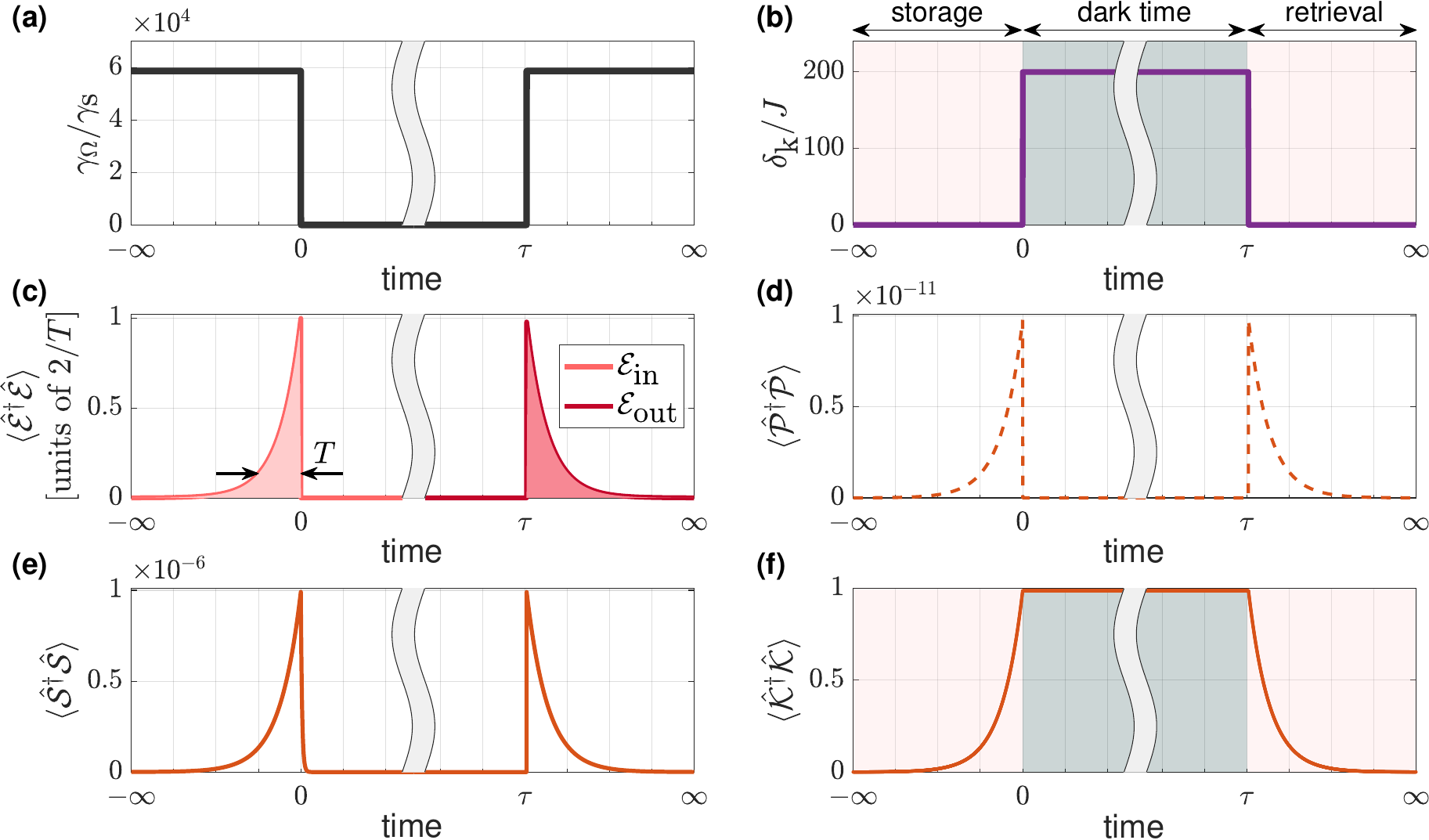}
\par\end{centering}
\centering{}\caption{{Example of storage and retrieval of a long pulse using the adiabatic mapping protocol. We numerically solve Eqs.\ (\ref{eq:Output field diff eq adiabatic K}-\ref{eq:dK_dt equation adiabatic}) for an exponentially-shaped signal field containing a single excitation [cf.~Eq.~(\ref{eq:exponentially shaped pulse input})], for $\Delta=\delta_{\textrm{s}}=0$ and for the square-profile control fields $\gamma_{\Omega}$ and $\delta_{\textrm{k}}$ in (a-b). Here $J=60\gamma_{\textnormal{s}}$, $\gamma_{\textnormal{s}}T=17$, and $C=100$.
The mapping of the input excitations $\langle\hat{\mathcal{E}}_{\textnormal{in}}^\dagger\hat{\mathcal{E}}_{\textnormal{in}}\rangle_{t}$ onto the atomic populations $\langle\hat{\mathcal{P}}^\dagger\hat{\mathcal{P}}\rangle_{t}$, $\langle\hat{\mathcal{S}}^\dagger\hat{\mathcal{S}}\rangle_{t}$,
and $\langle\hat{\mathcal{K}}^\dagger\hat{\mathcal{K}}\rangle_{t}$ is realized during storage, and the mapping onto $\langle\hat{\mathcal{E}}_{\textnormal{out}}^\dagger\hat{\mathcal{E}}_{\textnormal{out}}\rangle_{t}$ is realized during retrieval. The noble-gas spins adiabatically follow the input optical pulse, maintaining $\langle\hat{\mathcal{S}}^\dagger\hat{\mathcal{S}}\rangle_{t}\ll1$ throughout the protocol.
The optical pulse is retrieved by time reversing the storage sequence, yielding a total memory efficiency of $\eta_{\textrm{tot}}=0.98$.}
\label{fig:ADIABATIC}}
\end{figure*}

\begin{figure}[t]
\includegraphics[width=8.5cm]{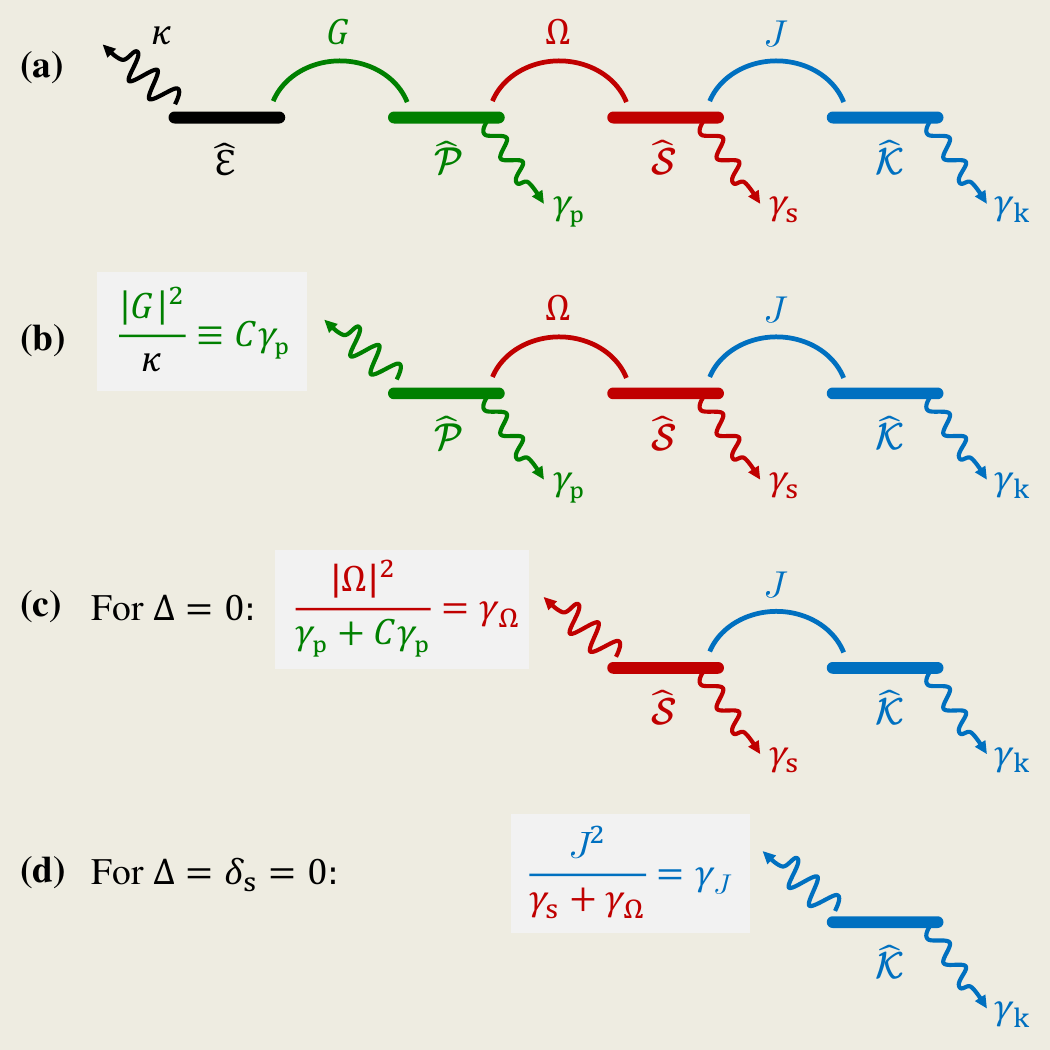}
\caption{Emergence of a decay rate $\gamma_{J}$ from the collective noble-gas spin $\hat{\mathcal{K}}$ during adiabatic retrieval. The picture is similar for adiabatic storage. (a) The optical cavity field $\hat{\mathcal{E}}$ couples to the collective optical dipole $\hat{\mathcal{P}}$, which couples to the  collective alkali spin $\hat{\mathcal{S}}$, which couples to the collective noble-gas spin $\hat{\mathcal{K}}$, with corresponding rates $G$, $\Omega$, and $J$. The optical field decays into the desired output field $\hat{\mathcal{E}}_{\text{out}}$ with rate $\kappa$. The rates $\gamma_\text{p}$, $\gamma_\text{s}$, and $\gamma_{\text{k}}$ encompass both relaxation and coupling to other (undesired) modes (not shown). (b) In the fast-cavity limit ($\kappa \gg G$), the cavity field adiabatically follows the optical dipole, giving rise to a decay rate $C\gamma_{\text{p}}$ from the optical dipole into the output field. This ``good'' decay rate $C\gamma_{\text{p}}$ competes with ``bad'' decay rate $\gamma_\text{p}$, contributing a factor of $C \gamma_{\text{p}}/(C\gamma_{\text{p}}+\gamma_{\text{p}})$ to the retrieval efficiency. (c) For moderate control fields ($C\gamma_{\text{p}} \gg \Omega$), the optical dipole adiabatically follows the alkali spin, giving rise to a decay rate $\gamma_{\Omega}$ from the alkali spin. This ``good'' decay rate $\gamma_{\Omega}$ competes with ``bad'' decay rate $\gamma_\text{s}$, contributing a factor of $\gamma_{\Omega}/(\gamma_{\Omega}+\gamma_\text{s})$ to the retrieval efficiency. (d) For $\gamma_\Omega+\gamma_\text{s} \gg J$ (corresponding to $T \gg 1/J$, since $T \sim 1/\gamma_J$), the alkali spin adiabatically follows the noble-gas spin, giving rise to a decay rate $\gamma_{J}$ from the noble gas spin. This ``good'' decay rate $\gamma_{J}$ competes with ``bad'' decay rate $\gamma_\text{k}$, contributing a factor of $\gamma_{J}/(\gamma_{J}+\gamma_\text{k})$ to the retrieval efficiency.
\label{fig:SCHEME}}
\end{figure}

\subsection{Adiabatic mapping\label{sec:Adiabatic-storage-on}}
We now turn to analyze the adiabatic mapping, which is suitable for storage and retrieval  of long-pulses satisfying $T\gg1/J$. In this
mapping scheme,  $\hat{\mathcal{S}}$
is kept unexcited during the storage and retrieval to avoid the loss
of excitations by the alkali-spin relaxation {[}cf.~Eq.~(\ref{eq:retreival efficiency}){]}. The principle is that $\hat{\mathcal{S}}$ can mimic the dynamics
of $\hat{\mathcal{P}}$ in the adiabatic regime, thus serving as a
mediator whose excitation is negligible.
 
 Below we examine the protocol's performance analytically using $T'=0$. We derive the equations of motion in the adiabatic regime, show that a direct coherent mapping $\hat{\mathcal{E}}_{\text{in}}\rightarrow\hat{\mathcal{K}}(0)$ is established during storage
and $\hat{\mathcal{K}}(\tau)\rightarrow\hat{\mathcal{E}}_{\text{out}}$ during retrieval. We then calculate the memory efficiency for an exponentially shaped pulse.

\subsubsection{Adiabatic equations of motion}\label{subsection:adiabatic}

We consider the dynamics of the uniform mode $\hat{\mathcal{S}}\left(t\right)$
in the adiabatic regime for $\gamma_{\Omega}\gg J,\,|Q\Omega/\sqrt{T}|$. Eq.~(\ref{eq:Uniform_alkali_spin_mode_diff_equation})
can then be approximated by
\begin{equation}
\hat{\mathcal{S}}=-\frac{Q\Omega^{*}\hat{\mathcal{E}}_{\text{in}}+iJ\hat{\mathcal{K}}-\hat{F}_{\mathcal{S}}}{\Gamma_{\Omega}+\gamma_{\text{s}}+i\delta_{\text{s}}}.\label{eq:S uniform adiabatic regime}
\end{equation}
Here $\hat{\mathcal{S}}\left(t\right)$ adiabatically follows the
operators $\hat{\mathcal{K}},$ $\hat{\mathcal{E}}_{\text{in}}$,
and $\hat{F}_{\mathcal{S}}$, similarly to the role of $\hat{\mathcal{P}}$
in Eq.~(\ref{eq:P adiabatic}). The alkali-spin excitation is small
$\langle\hat{\mathcal{S}}^\dagger \hat{\mathcal{S}}\rangle\leq J^{2}/|\Gamma_{\Omega}|^{2}\ll1$, allowing
for high memory efficiency according to Eqs.~(\ref{eq:storage efficiency})
and (\ref{eq:retreival efficiency}). By substituting the adiabatic
solution (\ref{eq:S uniform adiabatic regime}) in Eqs.~(\ref{eq:output-light-P})-(\ref{eq:P adiabatic-1})
and in Eq.~(\ref{eq:Uniform_noble_spin_mode_diff_equation}), we obtain
\begin{align}
\hat{\mathcal{E}}_{\text{out}} & =\left(\alpha-ia_{J}Q\Omega/J\right)\hat{\mathcal{E}}_{\text{in}}+a_{J}(\Omega/\Omega^{*})\hat{\mathcal{K}}+\hat{F}_{\mathcal{E}},\label{eq:Output field diff eq adiabatic K}\\
\partial_{t}\hat{\mathcal{K}} & =-(\gamma_{\text{k}}+\Gamma_{J}+i\delta_{\text{k}})\hat{\mathcal{K}}+a_{J}\hat{\mathcal{E}}_{\text{in}}+\hat{F}_{\mathcal{K}}.\label{eq:dK_dt equation adiabatic}
\end{align}
Here we define the stimulated coupling rate to the noble-gas spins
as
\begin{equation}
\Gamma_{J}\left(t\right)\equiv\frac{J^{2}}{\Gamma_{\Omega}\left(t\right)+\gamma_{\text{s}}+i\delta_{\text{s}}\left(t\right)},\label{eq:GammaJ}
\end{equation}
as well as the parameter $a_{J}\equiv iQ\Omega^{*}\Gamma_{J}/J$.
The vacuum noise operators are $\hat{F}_{\mathcal{K}}=\hat{f}_{\mathcal{K}}-i\Gamma_{J}\hat{F}_{\mathcal{S}}/J$
and $\hat{F}_{\mathcal{E}}=\hat{f}_{\mathcal{E}}+ia_{J}\Omega\hat{F}_{\mathcal{S}}/(J\Omega^{*})$,
and the parameters $Q$ and $\alpha$ are defined in Eqs.~(\ref{eq:Q_definition})
and (\ref{eq:alpha_definition}). 

{To exemplify this protocol, we solve numerically Eqs.~(\ref{eq:Output field diff eq adiabatic K}-\ref{eq:dK_dt equation adiabatic}), set the noise operators to zero, and use the input field in Eq.~(\ref{eq:exponentially shaped pulse input}). Here we solve the storage and retrieval stages for constant $\Delta=\delta_{\rm s}=\delta_{\rm k}=0$, set $T'=0$, and use $\gamma_{\Omega}(t)$, which follows a square temporal profile as shown in Fig.~\ref{fig:ADIABATIC}(a). In Fig.~\ref{fig:ADIABATIC}(c-f), we present the calculated quantum excitations of the optical input and output fields, and the atomic excitations $\langle\hat{\mathcal{P}}^\dagger\hat{\mathcal{P}}\rangle$, $\langle\hat{\mathcal{S}}^\dagger\hat{\mathcal{S}}\rangle$,
and $\langle\hat{\mathcal{K}}^\dagger\hat{\mathcal{K}}\rangle$ using
 $J=60\gamma_{\textnormal{s}},\,\gamma_{\textnormal{s}}T=17$, and $C=100$. The retrieved output field is presented in Fig.~\ref{fig:fast_Storage}(c), and the overall calculated memory efficiency in this example is  $\eta_{\textrm{tot}}=0.98$. Interestingly, here the alkali spin remains unexcited, $\langle\hat{\mathcal{S}}^\dagger\hat{\mathcal{S}}\rangle\ll1$, suppressing loss of the signal via alkali relaxation.}

It is also insightful to discuss the rate $\Gamma_{J}$ associated with the emergent coupling between noble-gas spins  and light. The imaginary part $\text{im}(\Gamma_{J})$
constitutes the frequency shift due to the spin-exchange coupling to
the alkali spins, which vanishes when operating at $\Delta=\delta_{\text{s}}=0$.
The real part 
\begin{equation}
\gamma_{J}\equiv\text{re}(\Gamma_{J})\label{eq:gammaJ}
\end{equation}
constitutes the relaxation inherited from the alkali spins. The alkali
spins themselves experience a relaxation at a high rate $\gamma_{\Omega}+\gamma_{\text{s}}$,
composed of radiative ($\gamma_{\Omega}$) and non-radiative ($\gamma_{\text{s}}$)
losses; both are partially inherited by the noble gas and are accounted
for in $\gamma_{J}$. The emergence of $\gamma_{J}$ and its analogy to $\gamma_\Omega$ are illustrated in Fig.~\ref{fig:SCHEME} for the case of retrieval. The intuition is the same for the case of storage.

Equation (\ref{eq:dK_dt equation adiabatic}) is a linear stochastic
differential equation and, similarly to Eq.~(\ref{eq:S_vector solution}),
can be solved by
\begin{align}
\hat{\mathcal{K}}\left(t\right) & =\Upsilon_{t,-\infty}\hat{\mathcal{K}}(-\infty)+\int_{-\infty}^{t}h_{J}(t,s)\mathcal{\hat{E}}_{\text{in}}\left(s\right)ds+\hat{\mathcal{W}}_{\mathcal{K}}(t),\label{eq:K0_solution_storage}
\end{align}where $h_{J}(t,s)=\Upsilon_{t,s}a_{J}\left(s\right)$, and $\hat{\mathcal{W}}_{\mathcal{K}}(t)=\int_{-\infty}^{t}\Upsilon_{t,s}\hat{F}_{\mathcal{K}}(s)ds$,
and the evolution function from time $t'$ to time $t$ is
\begin{equation}\Upsilon_{t,t'}=e^{-\int_{t'}^{t}[\gamma_{\text{k}}+\Gamma_{J}(s)+i\delta_{\text{k}}(s)]ds}.\end{equation}
Below we estimate the efficiencies of the storage and retrieval stages.

\subsubsection{Storage: $\hat{\mathcal{E}}_{\text{in}}\rightarrow\hat{\mathcal{\boldsymbol{K}}}$}

The spins are initially unexcited $\langle\hat{\mathcal{K}}^\dagger\hat{\mathcal{K}}\rangle_{-\infty}=0$,
so the first term in Eq.~(\ref{eq:K0_solution_storage}) vanishes.
We therefore get
\begin{equation}
\hat{\mathcal{K}}(0)=\int_{-\infty}^{0}h_{J}(0,s)\hat{\mathcal{E}}_{\text{in}}\left(s\right)ds+\hat{\mathcal{W}}_{\mathcal{K}}(0).\label{eq:K_storage_memory}
\end{equation}The transfer function $h_{J}$ satisfies the inequality
\begin{equation}
\int_{-\infty}^{0}\frac{1}{\acute{\eta}(t)}|h_{J}(0,t)|^{2}dt\leq1, 
\label{eff_4_lev_Stor}\end{equation}where the weight factor $\acute{\eta}$ is given by
\begin{equation}
\acute{\eta}(t)=\frac{C}{C+1}\frac{\gamma_{\Omega}(t)}{\gamma_{\Omega}(t)+\gamma_{\text{s}}}\frac{\gamma_{J}(t)}{\gamma_{J}(t)+\gamma_{\text{k}}}.
\end{equation}

We now calculate the storage efficiency for the exponentially-shaped signal in Eq.~(\ref{eq:exponentially shaped pulse input}) for $T>\gamma_{\text{s}}/J^2$, using a constant control pulse with amplitude $\gamma_{\Omega}=J^2T-\gamma_{\text{s}}$ (corresponding also to $\gamma_{J}=1/T$), and assuming negligible spin relaxation of the noble gas ($\gamma_{\text{k}}=0$). The storage efficiency is then given by
\begin{equation}
\eta_{\text{in}}=\frac{C}{C+1}\left(1-\frac{\gamma_{\text{s}}}{J^2T}\right).\label{eq:storage_efficiency}
\end{equation}
Turning off the control beam and applying a large magnetic field {[}$\delta(B)\gg J${]} after storage decouples the two spin gases and lets the noble gas act as a quantum memory, free of alkali-induced relaxation.

\subsubsection{Retrieval: $\hat{\mathcal{\boldsymbol{K}}}\rightarrow\hat{\mathcal{E}}_{\text{out}}$}

During retrieval, starting from $t=\tau$, there is no input signal ($\langle\hat{\mathcal{E}}^\dagger _{\text{in}}\hat{\mathcal{E}}_{\text{in}}\rangle=0$), and so the
second term in Eq.~(\ref{eq:K0_solution_storage}) vanishes. The noble-gas spin excitations are then given by
\begin{equation}
\langle\hat{\mathcal{K}}^\dagger\hat{\mathcal{K}}\rangle_t=e^{-2\int_{\tau}^{t}[\gamma_{\text{k}}+\gamma_{J}(s)]ds}\langle\hat{\mathcal{K}}^\dagger\hat{\mathcal{K}}\rangle_{\tau},
\end{equation}and substitution in Eq.~(\ref{eq:Output field diff eq adiabatic K}) yields the total output photon number
\begin{equation}
\int_{\tau}^{\infty}\langle\hat{\mathcal{E}}^\dagger_{\text{out}}\hat{\mathcal{E}}_{\text{out}} \rangle_{t'} dt'=\int_{\tau}^{\infty}|a_{J}(t')|^{2}\langle\hat{\mathcal{K}}^\dagger \hat{\mathcal{K}}\rangle_{t'} \label{adiabatic_eff} dt'.
\end{equation}
Similarly to $\Lambda$-system storage, the temporal shape of the output mode depends on the control field via the term $a_{J}(t')$. Substituting Eq.~(\ref{adiabatic_eff}) in Eq.~(\ref{eq:retrieval efficiency definition}) yields the output efficiency
\begin{equation}
\eta_{\text{out}}=\frac{C}{C+1}\int_{0}^{y(\infty)}\frac{\gamma_{\Omega}(y)}{\gamma_{\Omega}(y)+\gamma_{\text{s}}}\frac{\gamma_{J}(y)}{\gamma_{J}(y)+\gamma_{\text{k}}}e^{-y}dy\label{eq:adiabatic_eff_final},
\end{equation}
where \begin{equation}
y(t)=2\int_{\tau}^{t}\bigl(\gamma_{J}\left(s\right)+\gamma_{\text{k}}\bigr)ds.
\end{equation}
For the target exponentially-shaped mode we consider in Eq.~(\ref{eq:exponentially shaped pulse output}) and for $\gamma_{\text{k}}=0$, setting $\gamma_{J}=1/T$ and $\gamma_{\Omega}=J^2T-\gamma_{\text{s}}$ enable retrieval of light with efficiency
\begin{equation}
\eta_{\text{out}}=\eta_{\text{in}}=\frac{C}{C+1}\left(1-\frac{\gamma_{\text{s}}}{J^2T}\right).\label{eq:ret_equal_storage_efficiency}
\end{equation}

\subsection{Comparison between the sequential and adiabatic mappings}\label{sec:comparison_seq_adi}

It is insightful to compare the performance and validity regimes of the two protocols in subsections \ref{sec:Fast-storage-on} and \ref{subsection:adiabatic}  for the exponentially-shaped input and output signal fields in the limit $\gamma_{\text{k}}T\lll1$ (\emph{i.e.}, we account for nonzero $\gamma_{\text{k}}$ only during the long memory time $\tau$ between storage and retrieval). 
For the sequential mapping scheme, the memory efficiency is determined by combining Eqs.~(\ref{eq:Total efficiency}), (\ref{eq:relaxation in the dark noble gas}),
(\ref{eq:Storage-efficiency-alkali}), (\ref{eq:sequential-memory-efficiency-stage-2}), and (\ref{eq:Ret-efficiency-alkali}) into  $\eta_{\text{tot}}=\eta_{\text{in}}^{(\mathcal{E}\rightarrow\mathcal{S})}\eta_{\text{in}}^{(\mathcal{S}\rightarrow\mathcal{K})}\eta_{\mathrm{dark}}\eta_{\text{out}}^{(\mathcal{K}\rightarrow\mathcal{S})}\eta_{\text{out}}^{(\mathcal{S}\rightarrow\mathcal{E})}$, yielding 
\begin{equation}
\eta_{\text{tot}}=\left(\frac{C}{C+1}\right)^{2}\left(\frac{1-\gamma_{\text{s}}T}{1+\gamma_{\text{s}}T}\right)e^{-\frac{\pi\gamma_{\text{s}}}{J}}e^{-2\gamma_{\text{k}}\tau}.\label{eq:efficiency of the fast storage scheme}
\end{equation}
For the adiabatic mapping scheme, the memory efficiency is determined by combining Eqs.~(\ref{eq:storage_efficiency}) and (\ref{eq:ret_equal_storage_efficiency}),
\begin{equation}
\eta_{\text{tot}}=\left(\frac{C}{C+1}\right)^{2}\left(1-\frac{\gamma_{\text{s}}}{J^2T}\right)^{2}e^{-2\gamma_{\text{k}}\tau}.\label{eq:memory efficiency adiabatic}
\end{equation}
The memory efficiencies of both schemes approach unity for large cooperativity ($C\gg1$) and slow alkali relaxation. The latter amounts to the condition $\gamma_{\textnormal{s}}\ll J,1/T$ in the sequential scheme, and to $\gamma_{\textnormal{s}}\ll J^2T$ in the adiabatic scheme. Notably, when the pulse is long ($T\gg \gamma_{\text{s}}/J^2$), the efficiency of the adiabatic protocol approaches $C^2/(C+1)^2$ which is the maximal attainable efficiency of optically accessible $\Lambda$-type quantum memories. For the sequential mapping, the maximal attainable efficiency is further reduced by the factor $\exp(-\pi\gamma_{\text{s}}/J)$. On the other hand, when the pulse is short ($\gamma_{\text{s}}T\ll1$), the memory efficiency of the sequential scheme becomes independent of the pulse duration (as long as $TC\gamma_{\textnormal{p}}\gg1$), while the efficiency of the adiabatic scheme decreases as $[1-\gamma_{\text{s}}/(J^2T)]^2$ for $JT\gg1$. Interestingly, the adiabatic protocol can reach high memory efficiency even when the alkali and noble-gas spins are weakly coupled and the alkali relaxation is significant $\gamma_{\text{s}}\gtrsim J$, provided that the signal bandwidth is sufficiently small.  

\section{Possible experimental configurations \label{sec:Stochastic-evolution}}

The proposed quantum memory can be realized under a variety of experimental
conditions. We consider two types of configurations, which differ
in the pressure range of the buffer gas enclosed inside the glass cell; here `buffer gas' includes
both noble gas isotopes with nonzero nuclear spins used as the memory medium, and possibly additional inert
gases such as N$_2$ or noble gases with no nuclear spin. 

\newcommand{\commentout}[1]{}

\begin{table*}
\centering{}%
\begin{tabular}{|c|c|c|c|c|c|c|c|c|c|c|c|}
\hline
\# & Pressure & Noble gas & Alkali & Temperature & Additional & Coating & $\gamma_{\mathrm{s}}$ {[}$2\pi$ Hz{]}\commentout{{[}1/s{]}} & $J$ {[}$2\pi$ Hz{]}\commentout{{[}1/s{]}} & $C$ & 1/$\gamma_{\text{k}}$ & Efficiency\tabularnewline
 &  &  &  & {[}$^{\circ}\text{C}${]} & buffer gas &  &  &  &  &   & $\eta_{\textrm{tot}}$ \tabularnewline
\hline
\hline
\textbf{(1)} & High & $\text{\ensuremath{^{3}}He}$, 2 atm & K & 230 & $\text{N}_{2}$, 30 Torr & - & 2.4\commentout{15} & 110\commentout{680} & 37 & 100 {[}h{]} & 95\% | adiabatic\tabularnewline
 &  &  &  &  &  &  &  &  &  &  & 88\% | sequential\tabularnewline
\hline
\textbf{(2)} & High & $\text{\ensuremath{^{129}}Xe}$, 7 Torr & Rb & 150 & $\text{N}_{2}$, 1500 Torr & - & 1000\commentout{6800} & 92\commentout{580} & 2.7 & 22 {[}s{]} & 74\% | adiabatic\tabularnewline
\hline
\textbf{(3)} & Low & $\text{\ensuremath{^{129}}Xe}$, 0.2 Torr & Cs & 70 & - & Paraffin & 13\commentout{85} & 2.4\commentout{15} & 15 & 540 {[}s{]} & 94\% | adiabatic\tabularnewline
\hline
\textbf{(4)} & Low & $\text{\ensuremath{^{129}}Xe}$, 0.2 Torr & Cs & 90 & - & Alkane & 8\commentout{50} & 4.6\commentout{29} & 69 & 140 {[}s{]} & 98.5\% | adiabatic\tabularnewline
\hline
\end{tabular}\caption{\textbf{Possible experimental configurations and memory efficiencies.
(1-2) High pressure configurations. }(1) Mixture of potassium, 
helium-3, and $\text{N}_{2}$ (the last for mitigating radiation trapping during optical pumping) in a spherical glass cell
with a $1\,\text{cm}$ radius. The noble-gas spin state potentially
lives for $1/\gamma_{\text{k}}=100$ hours, limited by the dipole-dipole
limit. This configuration is compatible with both the sequential and
adiabatic storage schemes. (2) Mixture of rubidium, xenon-129, and
$\text{N}_{2}$ (the last to increase the molecular breakdown rate) in a cubical
glass cell with a $1\,\text{cm}$ edge-length. Here $J<\gamma_{\text{s}}$
such that the scheme is only suitable for an adiabatic operation.
Here $\gamma_{\text{k}}$ is limited by collisions with the alkali
atoms. For the high-pressure configurations (1-2), mapping of the optical
signal on the alkali spins can be implemented via Faraday teleportation
in a double-pass configuration, which realizes a beamsplitter interaction. \textbf{(3-4) For the low-pressure
configurations,} we consider a Cs-$\text{\ensuremath{^{129}}Xe}$
mixture enclosed in a cylindrical cell with anti-relaxation coating
for the alkali spins, where $\gamma_{\text{k}}$ is limited by collisions
with the alkali atoms. For these configurations mapping of the optical
signal onto the orientation moment of the alkali spins can be implemented
using any standard mapping scheme, such as EIT of linearly polarized
light tuned to the $F=4\rightarrow F=3$ optical transition of the
D1 line \citep{Katz-storage-of-light-2018}.\textbf{ }(3) Paraffin
coating, which allows for $N_{\text{e}}=1000$ bounces before spin randomization.
(4) Alkane coating, which allows for $N_{\text{e}}=10^{5}$ wall bounces
before spin randomization.
\label{tab:experimental_configs}}
\end{table*}

{Configurations with high buffer-gas pressure benefit
from higher SEOP efficiency and rate, higher exchange rate $J$, and lower alkali destruction rate $\gamma_{\text{diff}}$ due to collisions with the walls {[}cf.~Eq.~(\ref{eq:effective_alkali_diffusion_relaxation_rate}){]}. At high pressures, however, each of the alkali D$_1$ or D$_2$ lines used for the optical excitation is pressure broadened and appears as a single optical line. This broadening reduces the optical depth of the alkali medium and impedes optical resolution of or control over individual hyperfine transitions. Consequently, memory operation based on standard $\Lambda$-system protocols (\emph{e.g.}, EIT or Raman absorption) in practice deviates from that described by the simple $\Lambda$-type model and includes other undesired processes, such as four-wave mixing \citep{Katz-storage-of-light-2018}, which compromise memory efficiency and fidelity. 
Nevertheless, Faraday interaction in a double-pass configuration \citep{multipass1,multipass2} provides for a beam-splitter-like Hamiltonian, which is free of four-wave
mixing. This type of coupling is therefore consistent with our $\Lambda$-type modeling for the alkali spins. Moreover, at high buffer-gas pressures, the effect of tensor polarizability becomes negligible \citep{Happer-Book,katz-2013}, rendering the Faraday interaction with the orientation moment dominant, as desired for this memory.}

Low-pressure configurations benefit from strong atom-photon interaction
and enable employment of standard storage protocols (e.g.~EIT and Raman absorption) which utilize optical access and control over the state of different hyperfine levels. As diffusion
is faster in these configurations, anti-relaxation coated cells should
be used to avoid alkali-spin relaxation by collisions with the cell
walls \citep{Polzik-single-photon,Novikova-review,Katz-storage-of-light-2018}, but these coatings limit the operation temperature.

{The configurations we consider feature high optical depths, which may alleviate the need for an optical cavity and enable implementations based on free-space propagation. While exact calculations of a free-space model extend beyond the scope of this work, our main conclusions based on the fast-cavity regime should also remain, similar to Ref.~\citep{Gorshkov2}, approximately valid for cavity-free configurations. To approximate such configurations, we use the scaling found in Ref.~\citep{Gorshkov2} and relate the effective cooperativity of the medium to the optical depth in free space via $C\approx \textrm{OD}/5.8$, where OD is the free-space optical depth associated with attenuation of the power of the signal beam}.

In the following two subsections,  we analyze some exemplary possible configurations, which are also
summarized in Table~\ref{tab:experimental_configs}. 

\subsection{High buffer-gas pressure\label{subsec:High-buffer-gas} }
\emph{Mixture of potassium and helium-3.---}{The first configuration
we consider consists of a K-$\text{\ensuremath{^{3}}He}$ mixture
enclosed in a spherical glass cell with a $1\,\text{cm}$ radius.
We consider a $\text{\ensuremath{^{3}}He}$ density of $n_{\text{b}}=5.4\times10^{19}\,\text{\ensuremath{\unit{cm^{-3}}}}$
(corresponding to 2~atm
at ambient conditions), an alkali-metal
density of $n_{\text{a}}=5.2\times10^{14}\,\text{cm\ensuremath{^{-3}}}$
(corresponding to the vapor pressure at a temperature of $230\,^{\circ}\text{C})$
and $30\,\text{Torr}$ of $\text{N}_{2}$ to mitigate radiation trapping. We estimate a relaxation rate of about
$\gamma_{\textnormal{s}}=2\pi\cdot 2.4\,\text{\ensuremath{\text{Hz}}}$
during memory operation dominated
by collisions with the background gas (coupling to cell walls is reduced due to the slow diffusion, where the longest living alkali diffusion mode relaxes after $0.5\,\text{s}$
due to wall coupling) \citep{Firstenberg-Weak-collisions,Happer-Book}. This configuration is similar to the experimental conditions recently reported in \citep{katz-spectro,shaham-strong-coupling}.}

{The high buffer gas pressure broadens the optical D$_1$ transition of the potassium, such that its full width at half maximum (FWHM) is about $27\,\text{GHz}$.}
{Despite the broadening, a large resonant optical depth of about $\textnormal{OD}\approx220$
is expected, owing to the elevated alkali density, which corresponds to about $C\approx37$
for cavity-free configurations. The initialization of the alkali spins relies on optical pumping using a pump field, which is on only before the storage and retrieval stages (when the alkali and the noble gas are decoupled). Due to the high OD, efficient optical pumping of the alkali spins across the cell requires high intensity of the pump light at a large frequency detuning from the optical transition  \citep{shaham-strong-coupling,katz-spectro}. The pump light ``burns through'' the cell, as atoms whose spin is optically-pumped cease to absorb pump photons \citep{Happer-Walker-RMP} (but not signal photons, which differ in polarization and frequency). We estimate that alkali spin polarization of about $p_{\text{a}}\geq95\%$ can be obtained using $160\,\text{mW}$
of circularly polarized pump at $770$ nm, detuned by $70\,\text{GHz}$
from the D$_1$ transition, with a beam waist of about $1\,\text{cm}$ (yielding a pumping rate of $\sim 2\pi\cdot 50\,\textrm{Hz}$).

The noble gas spins can be initialized prior to the experiment via SEOP hyperpolarization to about $p_{\text{b}}\gtrsim75\%$ 
at a SEOP rate of $\sim(10\,\text{hour})^{-1}$
\citep{Walker-RMP-2017}. This day-long initialization is required only once, before the first memory operation, and is not required for continuous operation of the memory thereafter. These parameters allow for a high coherent coupling rate of $J\approx2\pi\cdot 110\,\textrm{Hz}$.
 We estimate that the collisional, magnetic-like spin-exchange shifts induced by one gas on the other are $230\,\mu\text{G}$
for the noble gas and $22\,\text{mG}$
for the alkali spins (cf.~Appendix~\ref{sec:Heisenberg-Bloch-Langevin-equati}). The high $J/\gamma_\mathrm{s}$ enables high-efficiency memory operation, with $\eta_{\text{tot}}\approx 88\%$ 
for the sequential protocol and $\eta_{\text{tot}}\approx 95\%$ for the adiabatic protocol. The memory efficiency in this configuration is limited mainly by the optical cooperativity of the alkali atoms, which can be increased if needed by increasing vapor density and pump power or by employing an optical cavity. 

The Faraday interaction in a double-pass configuration should be employed to map the photons onto the orientation moment $\hat{\mathcal{S}}$ of the alkali spin via direct interaction with the electron spin \citep{Polzik-RMP-2010}.
We consider a $1.5\,\text{W}$
 linearly polarized control beam detuned by $76\,\text{GHz}$
and estimate a coherent coupling rate of about
$\gamma_{\Omega}=2\pi\cdot 1.6\,\textrm{kHz}$ and $\Omega=2\pi\cdot 27\,\textrm{MHz}$,
taking into account the attenuation of the control field due to the small fraction of unpolarized alkali spins. The rate $\gamma_{\Omega}$ also approximates the signal bandwidth that can be stored and retrieved efficiently using the sequential scheme.
To enjoy the potential $1/\gamma_{\text{k}}=100$ hours coherence time of the noble gas, limited by self dipole-dipole
interactions, we need to minimize magnetic-field inhomogeneities and decrease  alkali density, e.g.~by lowering the cell temperature 
\citep{Walker-RMP-2017,key-Happer-1988}. Such a configuration could therefore feature an extraordinary time-bandwidth product exceeding $10^9$.}

\smallskip{}

\emph{Mixture of rubidium and xenon-129.---}{The second configuration we consider contains a Rb-$\text{\ensuremath{^{129}}Xe}$ mixture enclosed in
a cubical glass cell with a $1\,\text{cm}$ edge-length. We consider
a $\text{\ensuremath{^{129}}Xe}$ density of $n_{\text{b}}=2.5\times10^{17}\,\text{\ensuremath{\unit{cm^{-3}}}}$
(7~Torr at ambient conditions), an alkali density of $n_{\text{a}}=10^{14}\,\text{cm\ensuremath{^{-3}}}$
(temperature $150\,^{\circ}\text{C})$, and $1500\,\text{Torr}$
of $\text{N}_{2}$ to mitigate radiation trapping, decrease the diffusion coefficient, and increase the molecular breakdown rate of short-lived XeRb molecules ($\lesssim 0.1\,\text{ns}$) \citep{Happer-1984,Nelson-2001,Walker2006}. The latter is important for improving the SEOP efficiency and for decreasing the decay rate of the noble-gas spins due to the molecular interaction \citep{Nelson-2001}.} {The optical depth of the alkali medium is expected to be $\textnormal{OD}\approx16$}.

The alkali spins are maintained at a constant spin-polarization of $p_{\text{a}}=90\%$ by continuous optical-pumping. This pumping can be realized using a $250\,\text{mW}$ of circularly polarized $795$ nm beam, detuned $36\,\text{GHz}$ from the D$_1$ transition, whose linewidth is about $36\,\text{GHz}$. In this configuration, diffusion to cell walls occurs at a rate of about 
$2\pi\cdot 1\,\textrm{Hz}$
and is thus negligible. The pumping light shifts the alkali-spin resonance frequency by $2\;\textrm{kHz}$ and adds to its relaxation, which sums up to $\gamma_{\textrm{s}}\approx 2\pi\cdot1\,\textrm{kHz}$. 

{The $\text{\ensuremath{^{129}}Xe}$ atoms
are polarized via SEOP to $p_{\text{b}}=65\%$, limited by 
relaxation due to collisions with the alkali, yielding
$\gamma_{\text{k}}=2\pi\cdot 7\,\textrm{mHz}$.
As in the previous configuration, it is possible to decrease $\gamma_{\text{k}}$ during the memory time by lowering the alkali density. This configuration features 
$J=2\pi\cdot 92\;\textrm{Hz}$
and collisional shifts of $\sim 3.5\,\text{mG}$ exerted by the alkali on the xenon and $\sim 2.7\,\text{mG}$ exerted by the xenon on the alkali.

In this configuration, $J<\gamma_{\text{s}}$, so that it is most suitable for the adiabatic protocol. A $1\,\text{W}$ control beam, detuned by $38\,\text{GHz}$ and of $1\,\textrm{cm}$ waist, would provide for 
$\Omega=2\pi\cdot 40\,\textrm{MHz}$ and $\gamma_{\Omega}=2\pi\cdot 16\,\textrm{kHz}$
taking into account depletion of the control by unpolarized atoms. For ultra low-bandwidth pulses, we expect a memory efficiency of $\eta_{\textrm{tot}}\approx 74\%$. }

\subsection{Low buffer-gas pressure}

For the low-pressure configurations, we consider a Cs-$\text{\ensuremath{^{129}}Xe}$
mixture enclosed in a cylindrical cell with anti-relaxation coating for the cesium spins.
The alkali spins decohere by binary collisions with other alkali atoms,
by three-body collisions with xenon atoms, and by interaction with
the cell walls. The three-body collisions are associated with the formation
of short-lived Cs-$\text{\ensuremath{^{129}}Xe}$ (Van-der-Waals)
molecules, which limit the attainable degree of polarization $p_{\text{a}}$
and significantly increase the destruction rate $\gamma_{\text{s}}$.
To mitigate this process, we decrease the molecular formation rate
by considering a relatively low density $n_{\text{b}}=7\times10^{15}\,\text{\ensuremath{\unit{cm^{-3}}}}$
of $\text{\ensuremath{^{129}}Xe}$ (0.2~Torr). {At this pressure,
the molecular formation rate per alkali atom $R_{3}=Zn_{\text{b}}^{2}\approx2\pi\cdot0.4\,\textrm{Hz}$
is comparable to the binary spin-exchange collision rate $R_{2}=k_{\text{SE}}n_{\text{b}}$,
where $Z=8.5\times10^{-32}\times 2\pi\,\textrm{Hz}\,\text{cm}{}^{6}$ \citep{Happer-1985}
and $k_{\text{SE}}=6.5\times10^{-16}\times 2\pi\, \textrm{Hz}\,\text{cm}^{3}$
\citep{Happer-Book}.}
The low gas pressure corresponds to a large diffusion coefficient of $160\,\text{cm}^2/\text{s}$, leading to a mean free path of $0.2\,\text{mm}$ and mean free time of about $0.9\,\mu\text{s}$. For $1\,\text{cm}$ cell, the time for reaching the wall is of the order of $2\,\text{ms}$, indicating the importance of alkali-wall interaction.
Here we consider two configurations
based on different wall coatings, all compatible with the adiabatic
mapping scheme as $J<\gamma_{\text{s}}$.

\smallskip{}

\emph{Paraffin coating.---} Paraffin coating allows for $N_{\text{e}}\ge1000$
bounces before spin randomization \citep{Romalis-2010}.  { We consider
a cylindrical cell of length $L=3\,\text{cm}$ and radius $r=1\,\text{cm}$. The alkali density is $n_{\text{a}}=2\times10^{12}\,\text{cm\ensuremath{^{-3}}}$
(temperature $70^{\circ}\text{C})$, and 
$\gamma_{\text{s}}=2\pi\cdot 13\,\textrm{Hz}$
due to collisions with the walls and the background xenon.
{The optical line is inhomogeneously Doppler broadened by about $400\,\text{MHz}$ (FWHM), which yields a resonant optical depth of $\textrm{OD}\approx 90$ along the axis of the cylinder, corresponding to $C\approx 15$ for a cavity-free apparatus. In inhomogeneously broadened configurations, the OD can potentially be further enhanced by optical techniques \citep{Lahad-2019}.}

To allow for operation in the adiabatic regime for a duration longer than the
alkali-spin lifetimes, an alkali-spin polarization of $p_{\text{a}}=90\%$
is maintained by continuous optical-pumping \citep{Chalupczak,Budker-Book,Romalis-2010,Firstenberg-2019-pumping}.
While standard optical pumping would typically increase the relaxation rate in the dark 
by a factor of $1/(1-p_{\text{a}})$, %
pumping of the lower hyperfine manifold combined with frequent spin-exchange collisions allows for efficient pumping of the spin in the upper hyperfine manifold at reduced spin decoherence.
{To apply this pumping scheme, the pumping rate has to be faster than the spin-exchange rate 
$\sim 2\pi\cdot 220\,\textrm{Hz}$.
 On-resonance optical pumping with low-power light of $\sim 1\,\text{mW}$ readily provides a pumping rate of 
$\sim 2\pi\cdot 2\,\textrm{kHz}$.
} Under these conditions, the noble-gas
spin polarization is maintained at $p_{\text{b}}=50\%$ via binary
and molecular spin-exchange collisions assuming a coherence time of
$\gamma_{\mathrm{k}}^{-1}=9$ minutes limited by collisions with alkali
atoms. These parameters yield 
$J=2\pi\cdot2.4\;\textrm{Hz}$,
{ and collisional spin-exchange shifts corresponding to a magnetic-like field of $\sim 120\,\mu\text{G}$ for the xenon and $\sim 100\,\mu\text{G}$ for the alkali}.

Here mapping of the optical signal onto the orientation moment of the alkali spins can
be implemented with any standard mapping scheme, such as EIT with linearly-polarized light tuned to the $F=4\rightarrow F=3$ optical transition
of the D$_1$ line \citep{Katz-storage-of-light-2018}.
{For a $2\,\text{mW}$ linearly-polarized control beam at $894$ nm, detuned by $1\,\text{GHz}$ from the D$_1$ atomic line, we expect 
$\Omega=2\pi\cdot 1\,\textrm{MHz}$ and $\gamma_{\Omega}\approx 2\pi\cdot210\,\textrm{Hz}=15\gamma_{\text{s}}$,
for which a memory efficiency of 
$\eta_{\text{tot}}\approx 94\%$ can be realized.}}%

\smallskip{}

\emph{Alkane coating.---} A configuration based on alkane coating
could allow $N_{\text{e}}=10^{5}$ wall bounces before spin randomization
\citep{Romalis-2010,Budker-1-MIN}. {We consider cesium vapor density
of $n_{\text{a}}=9\times10^{12}\,\text{cm\ensuremath{^{-3}}}$ (temperature
$90^{\circ}\text{C})$ in a narrow cylindrical cell with $L=2\,\text{cm}$
and $r=1\,\text{mm}$ to mitigate radiation trapping.
{This aspect ratio gives an optical depth of 400 along the axis of the cell and only 26 across it.}
Maintaining
$p_{\text{a}}=90\%$ by continuous optical-pumping of the lower hyperfine
manifold { (using a $5.4\,\text{mW}$ pumping laser, for a pumping rate of 
$2\pi\cdot9\,\textrm{kHz}$,
 much faster than the 
$2\pi\cdot 1\,\textrm{kHz}$
spin-exchange rate)}, yields $p_{\text{b}}=50\%$, $\gamma_{\text{s}}=2\pi\cdot 8\,\textrm{Hz}$,

$\gamma_{\mathrm{k}}^{-1}=140\,\text{s}$, 
$J=2\pi\cdot 4.6\,\textrm{Hz}$,
and $C=69$ for a cavity-free configuration. Owing to the reduced alkali relaxation rate, here
$\gamma_{\Omega}=50\gamma_{\textnormal{s}}$
and 
$\Omega=2\pi\cdot 3.2\,\textrm{MHz}$
can be obtained with a $0.2\,\text{mW}$ control beam, detuned by $2.3\,\text{GHz}$ and of waist $1\,\text{mm}$. For this configuration, we estimate $\gamma_{J}\approx 2\pi\cdot 55\,\textrm{mHz}$ and a potential memory efficiency of
$\eta_{\text{tot}}\approx 98.5\%$ using the adiabatic protocol.}

\section{Noise characterization}\label{sec:Noise_characterization}
{Noise photons generated during the operation of a quantum memory might impact the overall fidelity. While the noise generated by the memory is independent of the state of the optical signal, its overall affect on the output light depends much on the particular state that is stored. For example, the effect of stray noise photons on weak coherent states or squeezed states of light (whose quantum correlations collectively reside in a large number of photons) would be less pronounced with respect to e.g.~a single photon. In this section, we consider potential noise sources in noble-gas-based memories and analyze their consequences.} 

\bigskip{}
\bigskip{}

{\subsection{Fluorescence noise} 
When alkali atoms are optically excited, they can return to the ground state while emitting a noise photon, either by spontaneous emission or by induced fluorescence due to collisions with buffer gas \cite{CIF1,CIF2}. Using ${\textrm{N}_2}$ as a quenching gas can considerably suppress emission of spontaneous photons, as it de-excites alkali atoms non-radiatively. Here we estimate the amount of noise photons arriving at the output mode due to flourescence in the absence of a quenching gas.

Assuming that $\gamma_{\mathrm{p}}$ is dominated by collisional broadening,  spontaneous excitation of alkali atoms is captured by the atomic noise operator $\hat{f}_{\mathcal{P}}$. Using Eq.~(\ref{eq:output light spins}), we find the relation between $\hat{f}_{\mathcal{P}}$ and the noise operator of the emitted noise photons: \begin{equation}\hat{f}_{\mathcal{E}}=\frac{\sqrt{2C\gamma_{\textnormal{p}}}}{\gamma_{\textnormal{p}}\left(1+C\right)+i\Delta}\hat{f}_{\mathcal{P}}.\label{eq:flor_noise}\end{equation} The operator $\hat{f}_{\mathcal{P}}$ is temporally white, and therefore the optical noise field at the detector $\hat{f}_{\mathcal{E}}$ has a spectral line-shape centered around the atomic line at $\omega_{\text{p}}$ and of width $\gamma_{\textnormal{p}}(1+C)$. Importantly, detuning the signal and control fields from resonance by $|\Delta|\gg\gamma_{\textnormal{p}}(1+C)$ renders the optical frequency of the noise photons different with respect to the signal photons by a frequency offset $\Delta$, which enables frequency filtering of the noise photons from the signal (e.g., by using optical filters).}

{To quantify the benefit of a large detuning $\Delta$, we estimate the fluorescence noise in the outgoing signal mode $\int_{T'+\tau}^{\infty}\langle\hat{\mathcal{E}}_{\textrm{out}}^\dagger\hat{\mathcal{E}}_{\textrm{out}}\rangle_{t} dt$ during the retrieval. The number of such noise photons due to fluorescence is determined by the contribution of $\hat{f}_{\mathcal{E}}$ in $\hat{\mathcal{E}}_{\textrm{out}}$ in Eq.~(\ref{eq:output light spins}), which is given by 
\begin{equation}
N_F=\int_{T'+\tau}^{\infty}\langle \hat{f}_{\mathcal{E}}^\dagger\hat{f}_{\mathcal{E}}\rangle_{t} dt\approx\frac{2C\gamma_{\textnormal{p}}}{\Delta^2}\int_{T'+\tau}^{\infty}\langle\hat{f}_{\mathcal{P}}^\dagger\hat{f}_{\mathcal{P}}\rangle_{t} dt.\end{equation} The last relation uses Eq.~(\ref{eq:flor_noise}) in the large detunning limit. Using Eq.~(\ref{noise_full}), we find that $\int\langle\hat{f}_{\mathcal{P}}^\dagger\hat{f}_{\mathcal{P}}\rangle_{t}dt\approx\gamma_\Omega$, and therefore $N_F\ll1$.

In summary, memories operating with off-resonant control can differentiate noise photons from signal photons by means of optical filtering, as well as by using molecular buffer gas which efficiently suppresses fluorescence via non-radiative channels.}

{\subsection{Imperfect alkali polarization} 
Imperfect alkali spin polarization can affect the memory performance. Here we consider a polarized medium satisfying  $\langle\sigma_{\downarrow\downarrow}\rangle\gg\langle\sigma_{\uparrow\uparrow}\rangle$. Imperfect optical pumping has two effects on the equations of motion.  First, an ensemble of thermally uncorrelated flipped spins with a density matrix $\sigma\equiv\sigma_1\otimes\sigma_2\otimes\ldots\otimes\sigma_{N_{\textrm{a}}}$ and degree of polarization $p_{\textrm{a}}=\left<\sigma_{\downarrow\downarrow}\right>< 1$ has a nonzero overlap with the collective mode of the desired spin-wave, thus initially populating that mode with {$\langle\hat{\mathcal{S}}^\dagger\hat{\mathcal{S}}\rangle_0=(1-p_{\text{a}})/2p_{\text{a}} \approx (1-p_{\text{a}})/2$} spin excitations \citep{Firstenberg-Weak-collisions}. Second, as the steady spin polarization is associated with dephasing, from the fluctuation-dissipation theorem, the quantum noise maintains steady  non-vacuum statistics. Consequently, the  normally ordered noise variance of the alkali spins increases to {$\langle\hat{f}_{\mathcal{S}}(t)^\dagger\hat{f}_{\mathcal{S}}(t')\rangle=\left[(1-p_{\textrm{a}})/p_{\textrm{a}}\right]\gamma_{\textrm{s}}\delta(t-t') \approx (1-p_{\textrm{a}})\gamma_{\textrm{s}}\delta(t-t')$}, 
while the anti-ordered variance is given by {$\langle\hat{f}_{\mathcal{S}}(t)\hat{f}_{\mathcal{S}}^\dagger(t')\rangle=\left[(1+p_{\textrm{a}})/p_{\textrm{a}}\right]\gamma_{\textrm{s}}\delta(t-t') \approx \left[2+(1-p_{\textrm{a}})\right]\gamma_{\textrm{s}}\delta(t-t')$}. 
This increased variance enters into the equations of motion of the two storage protocols. For the sequential storage, it enters via the noise process $\hat{\mathcal{W}}_{\mathcal{S}}$ in Eq.~(\ref{eq:S_vector solution}), where the detected noise photons in $\int_{T'+\tau}^{\infty}\langle\hat{\mathcal{E}}_{\textrm{out}}^\dagger\hat{\mathcal{E}}_{\textrm{out}}\rangle_{t} dt$ are given by $N=\int_{T'+\tau}^{\infty}|Q\Omega|^2\langle\hat{\mathcal{W}}_{\mathcal{S}}^\dagger\hat{\mathcal{W}}_{\mathcal{S}}\rangle_{t} dt$. Summing the contributions of the two effects, for high bandwidth pulses satisfying $\gamma_{\textrm{s}}T\ll1$ and $C\gg1$, we find that {$N\lesssim(1-p_{\textrm{a}})$}, which satisfies $N\ll1$  for a highly polarized medium. This noise mechanism is universal for all alkali-based quantum memories.} 

{\subsection{Imperfect noble-gas polarization}
Imperfect polarization $p_\text{b}<1$ of the noble-gas spins reduces the coupling rate $J\propto \sqrt{p_\text{b}}$ and increases the initial (incoherent) excitation of the collective uniform mode of the noble-gas ensemble by {$\langle\hat{\mathcal{K}}^\dagger\hat{\mathcal{K}}\rangle_0=(1-p_{\text{b}})/2p_{\text{b}}>0$}.
The latter could lead to readout and emission of a noise photon with probability {$(1-p_{\text{b}})/(1+p_{\text{b}}) \approx (1-p_{\text{b}})/2$},
 reducing the fidelity of the memory when storing single photons. However, this noise photon can potentially be read out prior to the storage process, via the same spin-exchange mechanism and retrieval procedure used for reading out the signal photon, leaving $\langle\hat{\mathcal{K}}^\dagger\hat{\mathcal{K}}\rangle_0=0$. Unlike the case of alkali spins, the rate at which the collective mode is replenished with incoherent excitations is $\gamma_{\text{k}}$, leaving ample time for the storage and retrieval of the signal photon. In fact, the read-out of the noise happens naturally in the sequential mapping process (strong-coupling regime $J\gg\gamma_\text{s}$), as the transfer process $\hat{\mathcal{S}}\rightarrow\hat{\mathcal{K}}$ is bi-directional and is accompanied by the reverse transfer $\hat{\mathcal{K}}\rightarrow\hat{\mathcal{S}}$. We shall explore these possibilities in future work.}

\section{Summary}\label{sec:Summary}

The analysis presented in this paper suggest that noble-gas spins and their interface via spin-exchange collisions can be used as an efficient interface to map non-classical light onto the collective and long-lived state of noble-gas spins, and therefore realize efficient quantum memories with potentially unprecedented time-bandwidth product. We outline various experimental configurations and mapping protocols which characterize and demonstrate the operation and performance of such memories.

The presented model is not limited to noble-gas spin systems and could potentially
be applied to analyze quantum memories in other four-level systems. These include, for example, an atomic system with a ladder of excited
electronic orbitals \citep{FLAME1,FLAME2,FLAME3}, as well as other hybrid systems
with both optically accessible and inaccessible spins, such as quantum
dots, diamond color-centers, and rare-earth impurities interacting
with nearby nuclear spins in the crystal. 

\begin{acknowledgments}
O.K., R.S., E.R., and O.F.\ acknowledge financial support by the Israel Science Foundation,
the European Research Council starting investigator grant Q-PHOTONICS
678674, the Pazy Foundation, the Minerva Foundation with funding from
the Federal German Ministry for Education and Research, and the Laboratory
in Memory of Leon and Blacky Broder. A.V.G.\ acknowledges support by ARL CDQI, ARO MURI, NSF PFC at JQI, AFOSR, AFOSR MURI, U.S.~Department of Energy Award No.~DE-SC0019449, DoE ASCR Accelerated Research in Quantum Computing program (award No.~DE-SC0020312), DoE ASCR Quantum Testbed Pathfinder program (award No.~DE-SC0019040), NSF PFCQC program, DARPA SAVaNT ADVENT, DoE QSA, and NSF QLCI (award No.~OMA-2120757).
\end{acknowledgments}

\appendix

\section{\label{sec:Heisenberg-Bloch-Langevin-equati}Heisenberg-Bloch-Langevin
equations}

The explicit form of the Heisenberg-Bloch-Langevin equations for $\hat{\sigma}_{\mu\nu}(\mathbf{r},t)$
is obtained by substituting $\mathcal{H}$ from Eqs.~(\ref{eq:Hamiltonian_tot})
and (\ref{eq:Hamiltonian_tot-1}) into Eq.~(\ref{eq:Heisenberg_Langevin_general_form}),
yielding
\begin{align}
\partial_{t}\hat{\sigma}_{\downarrow\textnormal{p}} & =-[\gamma_{\textnormal{p}}+i\Delta-i\tfrac{\left[I\right]}{4}\zeta(\hat{\sigma}_{\Downarrow\Downarrow}-\hat{\sigma}_{\Uparrow\Uparrow})]\hat{\sigma}_{\downarrow\textnormal{p}}\nonumber \\
+ & i\Omega(\mathbf{r},t)\hat{\sigma}_{\downarrow\uparrow}+ig(\mathbf{r})(\hat{\sigma}_{\downarrow\downarrow}-\hat{\sigma}_{\textnormal{pp}})\hat{\mathcal{E}}\label{eq:Bloch_no_approx_1}\\
+ & i\tfrac{\sqrt{\left[I\right]}}{2}\zeta\hat{\sigma}_{\uparrow\textnormal{p}}\hat{\sigma}_{\Downarrow\Uparrow}+D_{\text{a}}\nabla^{2}\hat{\sigma}_{\downarrow\textnormal{p}}+\hat{f}_{\downarrow\textnormal{p}},\nonumber \\
\nonumber \\
\partial_{t}\hat{\sigma}_{\downarrow\uparrow} & =-[\gamma_{\textnormal{s}}+i\delta_{\textnormal{s}}+i\tfrac{q_{I}-1}{4}\left[I\right]\zeta(\sigma_{\Downarrow\Downarrow}-\sigma_{\Uparrow\Uparrow})]\hat{\sigma}_{\downarrow\uparrow}\nonumber \\
+ & i\Omega^{*}(\mathbf{r},t)\hat{\sigma}_{\downarrow\textnormal{p}}-i\tfrac{\sqrt{\left[I\right]}}{2}\zeta(\hat{\sigma}_{\downarrow\downarrow}-\hat{\sigma}_{\uparrow\uparrow})\hat{\sigma}_{\Downarrow\Uparrow}\label{eq:Bloch_no_approx_2}\\
- & ig(\mathbf{r})\hat{\sigma}_{\textnormal{p}\uparrow}\hat{\mathcal{E}}+D_{\text{a}}\nabla^{2}\hat{\sigma}_{\downarrow\uparrow}+\hat{f}_{\downarrow\uparrow},\nonumber \\
\nonumber \\
\partial_{t}\hat{\sigma}_{\Downarrow\Uparrow} & =-[\gamma_{\textnormal{k}}+i\delta_{\textnormal{k}}-i\tfrac{\left[I\right]}{2}\zeta(\hat{\sigma}_{\downarrow\downarrow}+q_{I}\hat{\sigma}_{\uparrow\uparrow})]\hat{\sigma}_{\Downarrow\Uparrow}\label{eq:Bloch_no_approx_3}\\
- & i\tfrac{\sqrt{\left[I\right]}}{2}\zeta\hat{\sigma}_{\downarrow\uparrow}(\hat{\sigma}_{\Downarrow\Downarrow}-\hat{\sigma}_{\Uparrow\Uparrow})+D_{\text{b}}\nabla^{2}\hat{\sigma}_{\Downarrow\Uparrow}+\hat{f}_{\Downarrow\Uparrow},\nonumber
\end{align}
where $\gamma_{\textnormal{p}}\equiv\gamma_{\downarrow\textnormal{p}}$,~
$\gamma_{\textnormal{s}}\equiv\gamma_{\downarrow\uparrow}$, and $\gamma_{\textnormal{k}}\equiv\gamma_{\Downarrow\Uparrow}$.
These operators satisfy the commutation relations of continuous spin
operators
\begin{align}
\left[\hat{\sigma}_{\mu\nu}\left(\mathbf{r},t\right),\hat{\sigma}_{\alpha\beta}\left(\mathbf{r}',t\right)\right] & =\\
=\delta\left(\mathbf{r}-\mathbf{r}'\right) & \left(\delta_{\alpha\nu}\hat{\sigma}_{\mu\beta}\left(\mathbf{r},t\right)-\delta_{\beta\mu}\hat{\sigma}_{\alpha\nu}\left(\mathbf{r},t\right)\right).\nonumber
\end{align}
The equations can be further simplified by using the following assumptions:
the excited state is unpopulated, $\hat{\sigma}_{\textnormal{pp}}\approx0$
if the control power is kept low ($\Omega\ll\gamma_{\downarrow\textnormal{p}}$);
$\hat{\sigma}_{\textnormal{p}\uparrow}\approx0$ for weak input pulses
$\langle\hat{\mathcal{E}}^\dagger\hat{\mathcal{E}}\rangle\ll(\Omega/g)^2$; the collective operator $\hat{\sigma}_{\downarrow\downarrow}\approx p_{\text{a}}n_{\text{a}}$
is determined by the density of alkali atoms $n_{\text{a}}$ and by
the degree of ground-state polarization $p_{\text{a}}$, which is
kept high via optical pumping $p_{\text{a}}\rightarrow1$, such that
$\hat{\sigma}_{\uparrow\uparrow}\approx0$; and similarly the collective
operator $\hat{\sigma}_{\Downarrow\Downarrow}\approx p_{\text{b}}n_{\text{b}}$
is determined by the density of noble-gas atoms $n_{\text{b}}$ and
by the degree of polarization $p_{\text{b}}$, which satisfies $p_{\text{b}}\lesssim1$
owing to spin-exchange optical pumping (SEOP). We also note that the
collisional shift and the diffusion terms have negligible effect on
the optical linewidth. The simplified equations of motion are then
given by
\begin{align}
\partial_{t}\hat{\sigma}_{\downarrow\textnormal{p}} & =-(\gamma_{\textnormal{p}}+i\Delta)\hat{\sigma}_{\downarrow\textnormal{p}}+i\Omega(\mathbf{r},t)\hat{\sigma}_{\downarrow\uparrow}\label{eq:Bloch_equation_ge}\\
+ & iG(\vec{r})\hat{\mathcal{E}}+\hat{f}_{\downarrow\textnormal{p}},\nonumber \\
\partial_{t}\hat{\sigma}_{\downarrow\uparrow} & =-(\gamma_{\text{s}}+i\delta_{\text{s}}-D_{\text{a}}\nabla^{2})\hat{\sigma}_{\downarrow\uparrow}+i\Omega^{*}(\mathbf{r},t)\hat{\sigma}_{\downarrow\textnormal{p}}\label{eq:Bloch_equation_gs}\\
- & i(\zeta\sqrt{\left[I\right]}p_{\text{a}}n_{\text{a}}/2)\hat{\sigma}_{\Downarrow\Uparrow}+\hat{f}_{\downarrow\uparrow},\nonumber \\
\partial_{t}\hat{\sigma}_{\Downarrow\Uparrow} & =-(\gamma_{\text{k}}+i\delta_{\text{k}}-D_{\text{b}}\nabla^{2})\hat{\sigma}_{\Downarrow\Uparrow}\label{eq:Bloch_equation_mk}\\
- & i(\zeta\sqrt{\left[I\right]}p_{\text{b}}n_{\text{b}}/2)\hat{\sigma}_{\downarrow\uparrow}+\hat{f}_{\Downarrow\Uparrow},\nonumber
\end{align}
where $\Delta\rightarrow\Delta-\tfrac{\left[I\right]}{4}\zeta p_{\text{b}}n_{\text{b}}$.
The modified detuning of the alkali spins $\delta_{\text{s}}=\tilde{\delta}_{\text{s}}+(q_{I}-1)\left[I\right]\zeta n_{\text{b}}p_{\text{b}}/4$
accounts for the collisional shift that the alkali spins experience
due to the magnetized noble-gas spins. Similarly, the modified ~ detuning
of the noble-gas spins $\delta_{\text{k}}=\tilde{\delta}_{\text{k}}-\left[I\right]\zeta p_{\text{a}}n_{\text{a}}/2$
accounts for the collisional shift that the noble-gas spins experience
due to the magnetized alkali spins.

\section{Properties of the Quantum Noise\label{sec:Properties-of-Quantum}}

In this appendix, we present the properties of the quantum noise operators.
In the Heisenberg-Langevin picture, the relaxation of the quantum
operators is accompanied by stochastic quantum noise \citep{Zoller-Quantum-Noise-1999}.
In our model, we assume that the noise operators $\hat{f}_{\mu\nu}(\mathbf{r},t)$
defined in Eq.~(\ref{eq:Heisenberg_Langevin_general_form}) are temporally
white, satisfying
\begin{equation}
\langle\hat{f}_{\mu\nu}(\mathbf{r},t)\rangle=0\label{eq:quantum_noise_average}
\end{equation}
with variance
\begin{align}
\label{noise_full}\langle\hat{f}_{\mu\nu}(\mathbf{r},t)\hat{f}_{\alpha\beta}(\mathbf{r}',t')\rangle & =C_{\mu\nu\alpha\beta}\left(\mathbf{r},\mathbf{r}'\right)\delta\left(t-t'\right)\\
+\delta_{\nu\alpha}(\gamma_{\mu\nu}+\gamma_{\nu\beta} & -\gamma_{\mu\beta})\sigma_{\mu\beta}\left(\mathbf{r},t\right)\delta\left(\mathbf{r}-\mathbf{r}'\right)\delta\left(t-t'\right),\nonumber
\end{align}
where $C_{\mu\nu\alpha\beta}\left(\mathbf{r},\mathbf{r}'\right)$
is the diffusion noise correlation function for operators $\hat{\sigma}_{\mu\nu},\hat{\sigma}_{\alpha\beta}$.
The noise operators are essential for preserving the commutation relations $\bigl[\hat{\sigma}_{\mu\nu}\left(\mathbf{r},t\right),\hat{\sigma}_{\alpha\beta}\left(\mathbf{r}',t\right)\bigr]$.

For polarized spins, in the Holstein-Primakkof approximation, the
operators $\hat{\mathcal{P}}(\mathbf{r},t)$, $\hat{\mathcal{S}}(\mathbf{r},t)$,
and $\hat{\mathcal{K}}(\mathbf{r},t)$ act as local bosonic annihilation
operators. The noise terms $\hat{f}_{\mathcal{P}}=\hat{f}_{\downarrow\textnormal{p}}/\sqrt{p_{\text{a}}n_{\text{a}}}$,  $\hat{f}_{\mathcal{S}}=\hat{f}_{\downarrow\uparrow}/\sqrt{p_{\text{a}}n_{\text{a}}}$, 
and $\hat{f}_{\mathcal{K}}(\mathbf{r},t)=\hat{f}_{\Downarrow\Uparrow}/\sqrt{p_{\text{b}}n_{\text{b}}}$,
appearing in Eqs.~(\ref{eq:P_r_t_equation})-(\ref{eq:K_r_t_equation}),
then act as vacuum noise operators satisfying
\begin{equation}
\langle\hat{f}_{\text{q}}(\mathbf{r},t)\rangle=\langle\hat{f}_{\text{q}}^{\dagger}(\mathbf{r},t)\hat{f}_{\text{q}}(\mathbf{r}',t')\rangle=0
\end{equation}
and
\begin{align}
[\hat{f}_{\text{q}}(\mathbf{r},t),\hat{f}_{\text{q}}^{\dagger}(\mathbf{r}',t')] & =\langle\hat{f}_{\text{q}}(\mathbf{r},t),\hat{f}_{\text{q}}^{\dagger}(\mathbf{r}',t')\rangle\nonumber \\
=2(\gamma_{\text{q}} & -D_{\mathrm{q}}\nabla^{2})\delta(\mathbf{r}-\mathbf{r}')\delta\left(t-t'\right).\label{eq:quantum_noise_variance}
\end{align}
Here $\text{q}\in\{\mathcal{P},\mathcal{S},\mathcal{K}\}$, with $\gamma_{\mathcal{P}}\equiv\gamma_{\textnormal{p}}$,
$\gamma_{\mathcal{S}}\equiv\gamma_{\text{s}}$, $\gamma_{\mathcal{K}}\equiv\gamma_{k}$,
$D_{\mathcal{P}}=D_{\mathcal{S}}\equiv D_{\text{a}}$, and $D_{\mathcal{K}}\equiv D_{\mathrm{b}}$.
The first term in Eq.~(\ref{eq:quantum_noise_variance}) describes
a spatially-white noise with variance $2\gamma_{\text{q}}$, which
is associated with the relaxation rate $\gamma_{\text{q}}$ via the fluctuation-dissipation relations.
The second term is the diffusion component of the noise correlation
function, independent of the other relaxation mechanisms incorporated
in $\gamma_{\text{q}}$ \cite{CollectiveSpinStatesThermalDynamics}. The diffusion-induced decoherence rate of $\hat{\mathcal{P}}$ is negligible compared to $\gamma_{\text{p}}$ and so is the contribution of diffusion to the excited state noise.

\section{Spatial modes representation\label{sec:Spatial-modes-representation}}

Here we present the decomposition of the spin operators into
spatial mode functions and various choices for these functions. Equations
(\ref{eq:S_r_t_equation}) and (\ref{eq:K_r_t_equation}) contain
nonlocal terms due to atomic diffusion. Consequently, the evolution
of the spin operators is better described using a decomposition into
multiple nonlocal (spatial) modes. We therefore write $\hat{\mathcal{P}}(\mathbf{r},t)=\sum_{i}u_{i}^{(\text{p})}\left(\mathbf{r}\right)\hat{\mathcal{P}}_{i}\left(t\right)$,
$\hat{\mathcal{S}}\left(\mathbf{r},t\right)=\sum_{m}u_{m}^{(\text{s})}\left(\mathbf{r}\right)\hat{\mathcal{S}}_{m}\left(t\right)$,
and $\hat{\mathcal{K}}\left(\mathbf{r},t\right)=\sum_{n}u_{n}^{(\text{k})}\left(\mathbf{r}\right)\hat{\mathcal{K}}_{n}\left(t\right)$,
where each set of mode functions ($u_{\text{p}}$,$u_{\mathrm{s}}$,$u_{\text{k}}$)
is complete \citep{Polzik-RMP-2010}.

\subsection{Optical dipole $\hat{\mathcal{P}}$}

The optical dipole component that interacts with the field of the
cavity in Eq.~(\ref{eq:Output_field_cavity}) is defined by the spatial
overlap with the signal
\begin{equation}
\hat{\mathcal{P}}(t)=\sqrt{\frac{V_{\text{cav}}}{V}}\int_{V}f_{\varepsilon}^{*}\left(\mathbf{r}\right)\hat{\mathcal{P}}\left(\mathbf{r},t\right)d^{3}\mathbf{r}.\label{eq:P_t_definition}
\end{equation}
It is therefore fruitful to choose the set of modes $u_{i}^{(\text{p})}\left(\mathbf{r}\right)$
in which a specific mode $u_{0}^{(\text{p})}\left(\mathbf{r}\right)$
maximizes that integral, and all other modes are orthogonal. We thus
choose $u_{0}^{(\text{p})}\left(\mathbf{\mathbf{r}\in V}\right)=\sqrt{V_{\text{cav}}/V}f_{\varepsilon}\left(\mathbf{r}\right)$
{[}and $u_{0}^{(\text{p})}\left(\mathbf{\mathbf{r}\notin V}\right)=0${]}
and obtain $\hat{\mathcal{P}}\left(t\right)$ by substituting Eq.~(\ref{eq:P_local_adiabatic})
into Eq.~(\ref{eq:P_t_definition}),
\begin{equation}
\hat{\mathcal{P}}\left(t\right)=i\frac{\Omega(t)\sum_{m}b_{m}\hat{\mathcal{S}}_{m}\left(t\right)+\sqrt{2\gamma_{\textnormal{p}}C}\hat{\mathcal{E}}_{\text{in}}\left(t\right)-i\hat{f}_{\mathcal{P}}}{\gamma_{\textnormal{p}}\left(C+1\right)+i\Delta}.\label{eq:P adiabatic}
\end{equation}
The optical mode-matching parameter $b_{m}$ is given by
\begin{equation}
b_{m}=\sqrt{V_{\text{cav}}}\int_{V}u_{m}^{(\text{s})}\left(\mathbf{r}\right)u_{0}^{(\text{p})*}\left(\mathbf{r}\right)f_{\text{c}}(\mathbf{r})d^{3}\mathbf{r},\label{eq: b_m coefficients}
\end{equation}
characterizing the spatial overlap of the alkali-spin modes $u_{m}^{(\text{s})}\left(\mathbf{r}\right)$
with the mode $u_{0}^{(\text{p})}\left(\mathbf{r}\right)$ of the optical dipole,
weighted by the mode function of the optical control field in the
cavity $\sqrt{V_{\text{cav}}}f_{\text{c}}(\mathbf{r})$. The condition
$\sum_{m}|b_{m}|^{2}\leq1$ is satisfied. We also define the noise
operator of the single excited mode by $\hat{f}_{\mathcal{P}}(t)=\int_{V}u_{0}^{(\text{p})*}\left(\mathbf{r}\right)\hat{f}_{\mathcal{P}}(\mathbf{r},t)d^{3}\mathbf{r}$.

Correspondingly, Eq.~(\ref{eq:Output_field_cavity}) is transformed to\begin{equation}
\hat{\mathcal{E}}_{\text{out}}\left(t\right)=\alpha\hat{\mathcal{E}}_{\text{in}}-\sum_{m}p_{\text{out}}^{(m)}\left(t\right)\hat{\mathcal{S}}_{m}+\hat{f}_{\mathcal{E}},\label{eq:output light spins}
\end{equation}where we define
\begin{equation}
\alpha=\frac{\gamma_{\textnormal{p}}\left(1-C\right)+i\Delta}{\gamma_{\textnormal{p}}(1+C)+i\Delta},\label{eq:alpha_definition}
\end{equation}the coefficients $p_{\text{out}}^{(m)}\left(t\right)=Qb_{m}\Omega\left(t\right)$, and the additional noise operator for the output field $\hat{f}_{\mathcal{E}}=Q\hat{f}_{\mathcal{P}}$, where $Q$ is defined in Eq.~(\ref{eq:Q_definition}). 
\subsection{Noble-gas spin $\hat{\mathcal{K}}$}

The natural choice of mode functions for the collective noble-gas
spin is the set of eigenmodes of the diffusion-relaxation operator
\citep{CollectiveSpinStatesThermalDynamics}
\begin{equation}
\bigl(\gamma_{\textnormal{k}}-D_{\text{b}}\nabla^{2}\bigr)u_{n}^{(\text{k})}\left(\mathbf{r}\right)=\gamma_{n}^{\text{\ensuremath{(\text{k})}}}u_{n}^{(\text{k})}\left(\mathbf{r}\right),\label{eq:noble-gas-diffusion-relaxation-equation}
\end{equation}
assuming nondestructive (Neumman) boundary conditions. Here $\gamma_{n}^{\text{\ensuremath{(\text{k})}}}$
represents the relaxation rate of the $n^{th}$ mode. Using these
mode functions, the equations of motion of the noble-gas spin can
be written as
\begin{align}
\partial_{t}\hat{\mathcal{K}}_{n} & =-(\gamma_{n}^{\text{\ensuremath{(\text{k})}}}+i\delta_{\text{k}})\hat{\mathcal{K}}_{n}-iJ\sum_{m}c_{mn}^{*}\hat{\mathcal{S}}_{m}+\hat{f}_{\mathcal{K}}^{(n)},\label{eq:K_m mode}
\end{align}
where
\begin{equation}
c_{mn}=\int_{V}u_{m}^{(\text{s})*}\left(\mathbf{r}\right)u_{n}^{(\text{k})}\left(\mathbf{r}\right)d^{3}\mathbf{r}\label{eq: c coefficient  diffusion modes}
\end{equation}
describes the matching of the noble-gas spin modes to the alkali-spin
modes \citep{Firstenberg-Weak-collisions}. The matrix $[c_{mn}]$
is unitary, satisfying $\sum_{n}c_{mn}^{*}c_{nj}=\delta_{mj}$. The
normalized noise operators of the spin modes are $\hat{f}_{\mathcal{K}}^{(n)}=\int_{V}u_{n}^{(\text{k})*}\left(\mathbf{r}\right)\hat{f}_{\mathcal{K}}(\mathbf{r},t)d^{3}\mathbf{r}$.
In particular, the $n=0$ mode is the uniform spin mode, $u_{0}^{(\text{k})}\left(\mathbf{r}\right)=1/\sqrt{V}$,
unaffected by diffusion and exhibiting a minimal decay at a rate $\gamma_{\textnormal{k}}=\gamma_{0}^{\text{\ensuremath{\left(\text{k}\right)}}}$.
This mode is utilized here as the single mode of the long-lived quantum
memory
\begin{equation}
\hat{\mathcal{K}}\left(t\right)\equiv\hat{\mathcal{K}}_{0}\left(t\right)=\frac{1}{\sqrt{V}}\int\hat{\mathcal{K}}\left(\mathbf{r},t\right)d^{3}\mathbf{r}.
\end{equation}

\subsection{Alkali spin $\hat{\mathcal{S}}$}

Before choosing a particular basis for the alkali spins, we first
write Eq.~(\ref{eq:S_r_t_equation}) for a general basis $u_{m}^{(\text{s})}$.
Using Eq.~(\ref{eq:P adiabatic}), we obtain
\begin{align}
\partial_{t}\hat{\mathcal{S}}_{m} & =-(\gamma_{\textnormal{s}}+i\delta_{\text{s}})\hat{\mathcal{S}}_{m}-\sum_{j}\left(\Gamma_{\Omega}b_{m}^{*}b_{j}+d_{mj}\right)\hat{\mathcal{S}}_{j}\nonumber \\
- & iJ\sum_{n}c_{mn}\hat{\mathcal{K}}_{n}-p_{\mathrm{in}}^{(m)}\left(t\right)\hat{\mathcal{E}}_{\text{in}}+\hat{F}_{\mathcal{S}}^{(m)},\label{eq:S_m dynamics- P adiabatic}
\end{align}
Generally, the modes $u_{m}^{(\text{s})}$ are coupled by the atomic
diffusion, as represented by the coefficients
\begin{equation}
d_{mj}=-D_{\text{a}}\int_{V}d^{3}\mathbf{r}u_{m}^{(\text{s})\ast}\left(\mathbf{r}\right)\nabla^{2}u_{j}^{(\text{s})}\left(\mathbf{r}\right).\label{eq:coupling-between-S-modes-by-diffusion-operator}
\end{equation}
The coefficients $p_{\mathrm{in}}^{(m)}\left(t\right)=Qb_{m}^{*}\Omega^{*}\left(t\right)$
describe the coupling of each alkali-spin mode to the input light
field. The normalized noise operators of the alkali spin are $\hat{F}_{\mathcal{S}}^{(m)}=\int_{V}u_{m}^{(\text{s})*}\left(\mathbf{r}\right)\hat{f}_{\mathcal{S}}(\mathbf{r},t)d^{3}\mathbf{r}+ip_{\mathrm{in}}^{(m)}\hat{f}_{\mathcal{P}}/\sqrt{2C\gamma_{\textnormal{p}}}$,
including the quantum noise associated with the control beam (light-induced relaxation).

To choose a mode-function basis for the alkali spins, we examine eigenvalues
of the matrix $[\gamma_{\textnormal{s}}\delta_{mj}+\gamma_{\Omega}b_{m}^{*}b_{j}+d_{mj}]$, 
which correspond to the relaxation rates of the modes $\gamma_{m}^{\text{\ensuremath{\left(\text{s}\right)}}}$.
It follows that a convenient choice of mode basis exists in two limiting
regimes: when the dynamics is dominated by diffusion (\textit{e.g.},~in
the dark $\gamma_{\Omega}\ll D_{\textnormal{a}}/V^{2/3}$) and $[d_{mj}]$
is diagonal, or when the dynamics is dominated by power broadening
$(\gamma_{\Omega}\gg D_{\textnormal{a}}/V^{2/3})$ and $[b_{m}^{*}b_{j}]$
is diagonal. Here we consider these two regimes.

\subsubsection{\emph{Diffusion-dominated regime}}

In the regime $\gamma_{\Omega}\ll D_{\textnormal{a}}/V^{2/3}$, the
diffusion dominates over power broadening\emph{.} The natural choice
of basis is the set of eigenmodes of the diffusion-relaxation operator
for the alkali-metal spins
\begin{equation}
(\gamma_{\textnormal{s}}-D_{\text{a}}\nabla^{2})u_{m}^{(\text{s})}\left(\mathbf{r}\right)=\gamma_{m}^{\text{\ensuremath{\left(\text{s}\right)}}}u_{m}^{(\text{s})}\left(\mathbf{r}\right),\label{eq:alkali-diffusion-equation}
\end{equation}
satisfying destructive (Dirichlet) or partially-destructive (Robin)
boundary condition, depending on the quality of the anti-relaxation
coating of the cell walls \citep{Happer-Book,CollectiveSpinStatesThermalDynamics}. This
set of mode functions best applies for the sequential mapping protocol
in section \ref{sec:Fast-storage-on} and Appendix \ref{sec:multi-mode-sequential}.
In Appendix \ref{sec:Diffusion-relaxation-rate}, we calculate the
values of $b_{m}$ and $c_{mn}$ for an uncoated spherical cell.

\subsubsection{\emph{Light-dominated regime}}

In the regime $\gamma_{\Omega}\gg D_{\textnormal{a}}/V^{2/3}$, power
broadening due to the control beam dominates over diffusion. It is
then possible to engineer the spatial profile of the control field
such that the $m=0$ spin mode becomes the uniform mode $u_{0}^{(\text{s})}\left(\mathbf{r}\right)=1/\sqrt{V}$,
and $b_{m}=\delta_{m0}$. This can be realized by maintaining the
term $f_{\text{c}}(\mathbf{r})f_{\varepsilon}^{*}(\mathbf{r})$ constant
within the atomic cell, such that the term $\Omega^{*}(\mathbf{r},t)\hat{\mathcal{P}}(\mathbf{r},t)$
appearing in Eq.~(\ref{eq:S_r_t_equation}) is spatially independent.
This in turn yields a uniform two-photon excitation of the alkali
spin ensemble.
\begin{figure}
\centering{}\includegraphics[width=8.6cm]{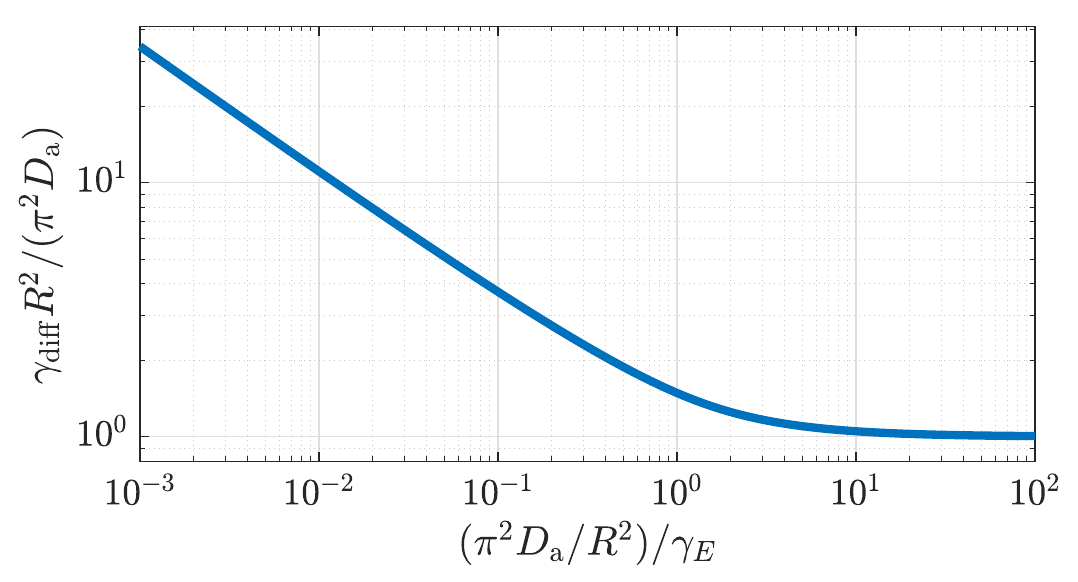}\caption{Effective diffusion-induced decay rate of the uniform mode of alkali spins as a function of the coupling duration $1/\gamma_{E}$.
\label{fig:Effective-decay-rate-of-alkali-uniform-distribution}}
\end{figure}

To exemplify this, in the large cavity limit, the spatial modes are
approximately the free-space modes $f_{\text{c}}(\mathbf{r})\approx e^{i\mathbf{k}_{\text{c}}\mathbf{r}}/\sqrt{V_{\text{cav}}}$
and $f_{\varepsilon}^{*}(\mathbf{r})=e^{-i\mathbf{k}_{\varepsilon}\mathbf{r}}/\sqrt{V_{\text{cav}}}$.
For an enclosure of length $L$, if $|\mathbf{k}_{\varepsilon}-\mathbf{k}_{\text{c}}|L\ll1$, 
then $f_{\text{c}}(\mathbf{r})f_{\text{\ensuremath{\varepsilon}}}^{*}(\mathbf{r})\approx1/V_{\text{cav}}$
and the input signal excites the uniform spin mode efficiently. This
condition is often satisfied in experiments when the signal and control
fields are nearly degenerate. Under these conditions, we get $b_{0}=c_{10}^{*}=1$
while $b_{m}=c_{m0}^{*}=0$ for $m\neq1$. We can
then approximate the dynamics in Eq.~(\ref{eq:Uniform_alkali_spin_mode_diff_equation})
with the use of the uniform spin operator
\begin{equation}
\hat{\mathcal{S}}\left(t\right)\equiv\frac{1}{\sqrt{V}}\int_{V}\hat{\mathcal{S}}\left(\mathbf{r},t\right)d^{3}\mathbf{r}.\label{eq: alkali uniform mode}
\end{equation}
To account for the contribution of higher spatial modes in Eq.~(\ref{eq:S_m dynamics- P adiabatic}),
we approximate the multi-exponential decay using a single effective
rate
\begin{equation}
\gamma_{\textnormal{s}}\rightarrow\gamma_{\textnormal{s}}+\gamma_{\text{diff}}.\label{eq:diffusion correction}
\end{equation}
While for the sequential scheme it is natural to decompose the uniform
spin mode into the diffusion eigenmodes, which lead to a multi-exponential
decay {[}cf.~Eqs.~(\ref{eq:S_m mode dark-1-2})-(\ref{eq:K_m mode dark-1-2}){]},
for the adiabatic scheme the dynamics can be well approximated by
a single exponential deacy, since $\gamma_{\Omega}$ dominates over
the diffusion rate of the least decaying modes. We therefore use Eq.~(\ref{eq:Uniform_alkali_spin_mode_diff_equation}),
which neglects the contribution of the other spatial modes to the
uniform alkali mode. To best approximate the dynamics and the diffusion-induced
relaxation in Eq.~(\ref{eq:diffusion correction}), we define the
effective rate
\begin{equation}
\gamma_{\text{diff}}=-\gamma_{E}\ln\biggl[\sum_{m=0}^{\infty}\left|c_{m0}\right|^{2}\exp\Bigl(\frac{\gamma_{\Omega}+\gamma_{\textnormal{s}}-\gamma_{m}^{\text{(s)}}}{\gamma_{E}}\Bigr)\biggr],\label{eq:effective_alkali_diffusion_relaxation_rate}
\end{equation}
which weighs the contribution of the different diffusion mode within
some coupling duration $1/\gamma_{E}$. In Fig.$\,$\ref{fig:Effective-decay-rate-of-alkali-uniform-distribution},
we present the diffusion decay rate $\gamma_{\text{diff}}$ in an uncoated cell with respect
to the decay rate $D_{\textnormal{a}}\pi^{2}/R^{2}$ of the lowest-order diffusion mode.
In the calculations presented in the main text, we choose $\gamma_{E}=\gamma_{\Omega}+\gamma_{\textnormal{s}}$,
such that $\gamma_{\text{diff}}$ approaches $D_{\textnormal{a}}\pi^{2}/R^{2}$.

\section{\label{sec:Diffusion-relaxation-rate} Numerical evaluation of the
spatial mode decomposition for a spherical cell}

\begin{figure}
\begin{centering}
\includegraphics[width=8.6cm]{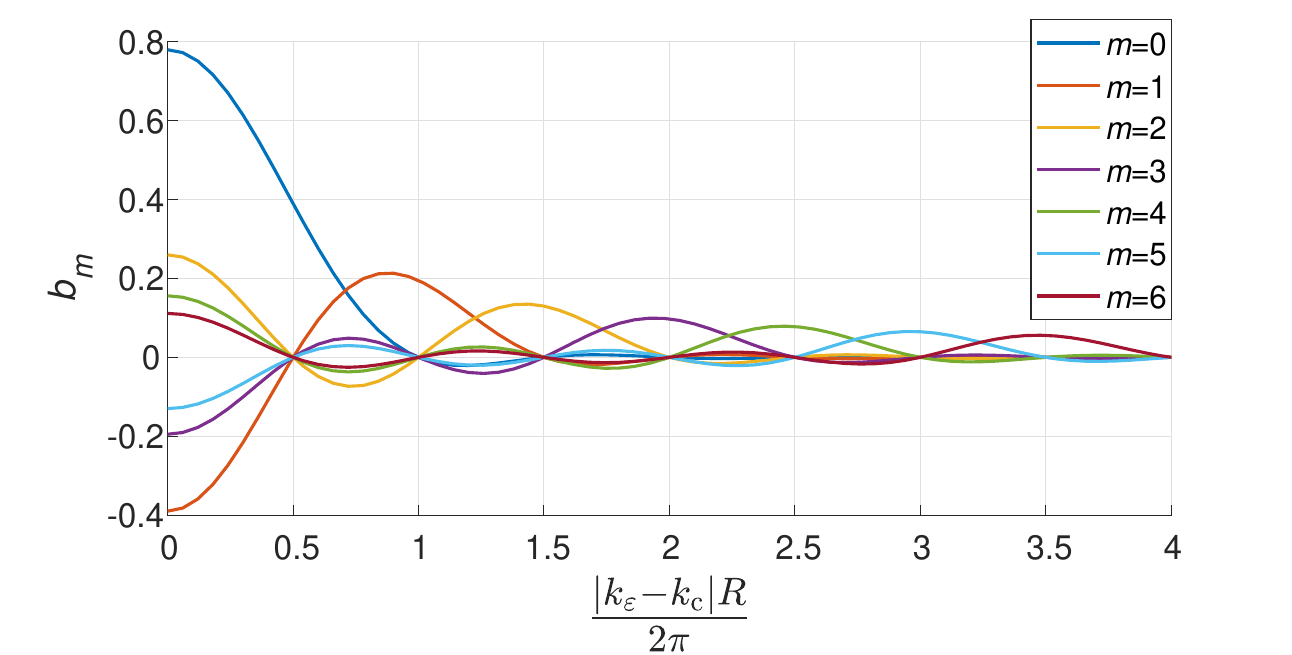}
\par\end{centering}
\caption{Mode overlap of the uniform spin distribution with the signal and
control fields in the cavity, cf.~Eq.$\,$(\ref{eq: b_m coefficients}),
for the first seven modes. The condition $|\boldsymbol{k}_{\varepsilon}-\boldsymbol{k}_{\textnormal{c}}|R\ll1$
is typically satisfied.\label{fig:bm_vs_dkR}}
\end{figure}

In this appendix, we present a numerical evaluation of the overlap
coefficients $b$ and $c$ and derive an approximate expression for
the diffusion relaxation of the uniform alkali-spin mode. We consider
a spherical cell of radius $R$ and define the diffusion modes following
Ref.~\citep{CollectiveSpinStatesThermalDynamics}. In a spherical cell, the diffusion
modes are labeled by $\left(m,\ell,\mu\right)$, where $m$ characterizes
the radial dependence, and $\left(\ell,\mu\right)$ represent the
angular symmetry. Since we ultimately consider storage on the noble-gas
uniform mode, we use only the spherically-symmetric modes, \emph{i.e.}
$\ell=\mu=0$. We therefore use a single label $m$ to index the modes.

\begin{table}
\begin{centering}
\begin{tabular}{|c|c|c|}
\hline
mode number $(m)$ & $\frac{\gamma_{m}^{\text{(s)}}-\gamma_{\text{s}}}{D_{\text{a}}\pi^{2}/R^{2}}$ & $\frac{\gamma_{m}^{\text{(k)}}-\gamma_{\text{k}}}{D_{\text{b}}\pi^{2}/R^{2}}$\tabularnewline
\hline
\hline
0 & 1 & 0\tabularnewline
\hline
1 & 4 & 2.05\tabularnewline
\hline
2 & 9 & 6.05\tabularnewline
\hline
3 & 16 & 12.05\tabularnewline
\hline
4 & 25 & 20.05\tabularnewline
\hline
5 & 36 & 30.05\tabularnewline
\hline
6 & 49 & 42.05\tabularnewline
\hline
\end{tabular}
\par\end{centering}
\caption{Diffusion-induced decay rates for the first seven spherically symmetric
modes of alkali spins ($\gamma_{m}^{\text{(s)}}-\gamma_{\text{s}}$)
and noble-gas spins ($\gamma_{m}^{\text{(k)}}-\gamma_{\text{k}}$)
in an uncoated spherical cell {[}cf.~Eqs.$\,$(\ref{eq:noble-gas-diffusion-relaxation-equation})
and (\ref{eq:alkali-diffusion-equation}){]}. \label{tab:diffusion-decay-rates}}
\end{table}

\subsection{Representation of $b$}

We calculate the overlap $b_{m}$
between the alkali spin and the optical field, as defined in Eq.$\,$(\ref{eq: b_m coefficients}).
We set the $\hat{\mathbf{z}}$ axis as the cavity axis, such that
for the control we have $f_{\textnormal{c}}\left(\mathbf{r}\right)=e^{-ik_{\textnormal{c}}z}/\sqrt{V_{\text{cav}}}$,
and the optical dipole spatially follows the signal field $u_{0}^{(\text{p})*}\left(\mathbf{r}\right)=e^{i\mathbf{k}_{\varepsilon}\cdot\mathbf{r}}/\sqrt{V}$.
With these, we get for the spherically-symmetric modes
\begin{align}
b_{m} & =\int_{0}^{R}r^{2}dr\int_{0}^{\pi}\sin\theta d\theta\int_{0}^{2\pi}b_{m}(r,\theta,\varphi)d\varphi,
\end{align}
where
\begin{equation}
b_{m}(r,\theta,\varphi)=\frac{\sqrt{3}j_{0}(\pi x_{m}r/R)\cdot e^{i(k_{\varepsilon}-k_{\textnormal{c}})z}}{A_{m}4\pi R^{3}}.
\end{equation}
Here $j_{0}\left(x\right)$ denotes the zeroth-order spherical Bessel
function, $\pi x_{m}$ is its $m^{\text{th}}$ root such that $j_{0}(\pi x_{m})=0$,
and $A_{m}$ is the normalization factor defined as
\begin{equation}
A_{m}^{2}R^{3}=\int_{0}^{R}r^{2}j_{0}^{2}\left(\pi x_{m}r/R\right)dr.
\end{equation}
Expanding the expression $\exp(i(k_{\varepsilon}-k_{\textnormal{c}})z)$
as a series of Bessel functions yields
\begin{align}
b_{m} & =\frac{\sqrt{3}}{A_{m}}\int_{0}^{1}\xi^{2}j_{0}(\pi x_{m}\xi)\biggl[J_{0}\bigl((k_{\varepsilon}-k_{\textnormal{c}})R\xi\bigr)\nonumber \\
+ & 2\sum_{\kappa=1}^{\infty}\frac{\left(-1\right)^{\kappa}J_{2\kappa}\bigl((k_{\varepsilon}-k_{\textnormal{c}})R\xi\bigr)}{1-4\kappa^{2}}\biggr]d\xi.
\end{align}
In Fig.$\,$\ref{fig:bm_vs_dkR}, we present $b_{m}$ for the first
seven modes $m=0,\ldots,6$. For standard alkali-spin memories, the
condition $\left|k_{\varepsilon}-k_{\textnormal{c}}\right|R\ll1$
is typically satisfied, yielding the identity $b_{m}=c_{m0}$.

\subsection{Representation of $c$}

In an uncoated cell, the alkali and noble-gas spins interact differently
with the surface of the glass wall, leading to different boundary
conditions for the diffusion of spins (Neumann for noble-gas spins and
Dirichlet for alkali spins) \cite{CollectiveSpinStatesThermalDynamics}. This in turn leads to different sets
of radial eigenmodes of the two spin species in a spherical cell.
Under these conditions, we calculate the diffusion-induced decay rates
$\gamma_{m}^{(\text{s})}-\gamma_{\text{s}}$ for the alkali spins
and $\gamma_{n}^{(\text{k})}-\gamma_{\text{k}}$ for the noble-gas
spins (see Table \ref{tab:diffusion-decay-rates}) and the overlap
coefficients for the modes of the two species $c_{mn}$ (see Table
\ref{tab:c-values-diffusion-modes-overlap}).

\begin{table}
\centering{}%
\begin{tabular}{|c|c|c|c|c|c|c|c|}
\hline
 & $n=0$ & $n=1$ & $n=2$ & $n=3$ & $n=4$ & $n=5$ & $n=6$\tabularnewline
\hline
\hline
$m=0$ & 0.780 & 0.609 & -0.126 & 0.058 & -0.033 & 0.022 & -0.016\tabularnewline
\hline
$m=1$ & -0.390 & 0.652 & 0.622 & -0.158 & 0.079 & -0.049 & 0.033\tabularnewline
\hline
$m=2$ & 0.260 & -0.275 & 0.0647 & 0.0627 & -0.173 & 0.091 & -0.058\tabularnewline
\hline
$m=3$ & -0.195 & 0.182 & -0.256 & 0.644 & 0.629 & -0.181 & 0.098\tabularnewline
\hline
$m=4$ & 0.156 & -0.139 & 0.168 & -0.246 & 0.643 & 0.631 & -0.187\tabularnewline
\hline
$m=5$ & -0.130 & 0.112 & -0.128 & 0.159 & -0.239 & 0.642 & 0.632\tabularnewline
\hline
$m=6$ & 0.111 & -0.095 & 0.104 & -0.121 & 0.154 & -0.235 & 0.641\tabularnewline
\hline
\end{tabular}\caption{The overlap coefficients $c_{mn}$ between the $m^{\text{th}}$ diffusion
eigenmode of alkali spins and the $n^{\text{th}}$ diffusion eigenmode
of noble-gas spins in an uncoated cell {[}cf.~Eq.$\,$(\ref{eq: c coefficient  diffusion modes}){]}.\label{tab:c-values-diffusion-modes-overlap}}
\end{table}

\section{\label{sec:conservation-of-excitations}Conservation of excitations}

Here we identify an integral relation, which can be viewed as a conservation
law for the excitations. The excitations in the optical signal $\hat{\mathcal{E}}_{\text{in}}$
can be exchanged between the spin operators $\hat{\mathcal{P}}$,
$\hat{\mathcal{S}}$, $\hat{\mathcal{K}}$, and finally be transferred
to $\hat{\mathcal{E}}_{\text{out}}$. Using Eqs.~(\ref{eq:P adiabatic-1}),
(\ref{eq:Uniform_alkali_spin_mode_diff_equation}), and (\ref{eq:Uniform_noble_spin_mode_diff_equation}),
we write the relation
\begin{align}
\langle\hat{\mathcal{E}}_{\text{out}}^\dagger\hat{\mathcal{E}}_{\text{out}}\rangle & -\langle\hat{\mathcal{E}}_{\text{in}}^\dagger \hat{\mathcal{E}}_{\text{in}}\rangle+\partial_{t}\bigl(\langle\hat{\mathcal{P}}^\dagger \hat{\mathcal{P}} \rangle+\langle\hat{\mathcal{S}}^\dagger\hat{\mathcal{S}}\rangle+\langle\hat{\mathcal{K}}^\dagger\hat{\mathcal{K}}\rangle\bigr)\nonumber \\
=-2\bigl(\gamma_{\textnormal{p}} & \langle\hat{\mathcal{P}}^\dagger\hat{\mathcal{P}}\rangle+\gamma_{\text{s}}\langle\hat{\mathcal{S}}^\dagger\hat{\mathcal{S}}\rangle+\gamma_{\text{k}}\langle \hat{\mathcal{K}}^\dagger \hat{\mathcal{K}}\rangle\bigr).
\end{align}
It is evident that, in a lossless cavity, excitations decay only through the relaxations
$\gamma_{\textnormal{p}}$, $\gamma_{\text{s}}$, and $\gamma_{\text{k}}$
during the time that $\hat{\mathcal{P}}$, $\hat{\mathcal{S}}$, and
$\hat{\mathcal{K}}$ are excited. Upon integration, we get the relation
\begin{align}
\int_{t_{1}}^{t_{2}} & \langle\hat{\mathcal{E}}_{\text{out}}^\dagger\hat{\mathcal{E}}_{\text{out}}-\hat{\mathcal{E}}_{\text{in}}^\dagger \hat{\mathcal{E}}_{\text{in}}\rangle_{t} dt+\langle\hat{\mathcal{P}}^\dagger\hat{\mathcal{P}}\rangle_{t_{2}}-\langle\hat{\mathcal{P}}^\dagger \hat{\mathcal{P}}\rangle_{t_{1}}\nonumber \\
+ & \langle\hat{\mathcal{S}}^\dagger\hat{\mathcal{S}}\rangle_{t_{2}}-\langle\hat{\mathcal{S}}^\dagger \hat{\mathcal{S}}\rangle_{t_{1}}+\langle\hat{\mathcal{K}}^\dagger \hat{\mathcal{K}}\rangle _{t_{2}}-\langle\hat{\mathcal{K}}^\dagger \hat{\mathcal{K}} \rangle_{t_{1}}\label{eq:Integral excitation conservation}\\
= & -2\int_{t_{1}}^{t_{2}}\Bigl(\gamma_{\textnormal{p}}\langle\hat{\mathcal{P}}^\dagger\hat{\mathcal{P}}\rangle_{t}+\gamma_{\text{s}}\langle\hat{\mathcal{S}}^\dagger\hat{\mathcal{S}}\rangle_{t}+\gamma_{\text{k}}\langle\hat{\mathcal{K}}^\dagger\hat{\mathcal{K}}\rangle_{t} \Bigr) dt,\nonumber
\end{align}
which describes the conservation of excitations.

\section{\label{sec:alkali_memory} Storage and Retrieval of alkali memories}

In this appendix, we review the formalism of Ref.~\citep{Gorshkov1}
to describe storage and retrieval of optical memories using alkali
spins.

Equation (\ref{eq:Uniform_alkali_spin_mode_diff_equation}) is a linear
stochastic differential equation for the alkali-spin operator. The solution for this equation is given by
\begin{align}
\mathcal{\hat{S}}\left(0\right) & =\Phi_{0,-\infty}\mathcal{\hat{S}}(-\infty)-\int_{-\infty}^{0}\Phi_{0,t}Q\Omega^{*}\hat{\mathcal{E}}_{\text{in}}\left(t\right)dt+\hat{\mathcal{W}}_{\mathcal{S}}\left(0\right).\label{eq:S_vector solution}
\end{align} The first term describes the deterministic evolution of $\mathcal{\hat{S}}$
in the absence of an input signal. The second term describes the response
of the spins to the input optical signal. The third term describes
the stochastic response of the spin via the stochastic quantum process
$\hat{\mathcal{W}}_{\mathcal{S}}(0)=\int_{-\infty}^{0}\Phi_{0,\tau}\hat{F}_{\mathcal{S}}\left(\tau\right)d\tau$.
The operator $\Phi_{0,t}$ is the evolution in the interval $t<T$,
given by
\begin{equation}
\Phi_{0,t}=\exp\left[-\int_{t}^{0}[\Gamma_{\Omega}(s)+\gamma_{\textnormal{s}}+i\delta_{\text{s}}]ds\right].
\end{equation}
Note that this solution accounts for the diffusion-induced relaxation
of the alkali spins in the power-broadened regime $\gamma_{\Omega}\gg D_{\text{a}}/V^{2/3}$
via Eqs.~(\ref{eq:diffusion correction}) and (\ref{eq:effective_alkali_diffusion_relaxation_rate}).
We now focus on the storage and retrieval stages for $T'=0$.

\smallskip{}

\emph{Storage}.--- Initially $\langle\mathcal{\hat{S}}^\dagger\mathcal{\hat{S}}\rangle_{(t=-\infty)}=0$, so
the first term in Eq.~(\ref{eq:S_vector solution}) vanishes. By
defining the transfer function
\begin{equation}
h_{\Omega}(0,t)=-Q\Omega^{*}\Phi_{0,t},\label{eq:f transfer function}
\end{equation}
we write the alkali spin after storage as
\begin{equation}
\mathcal{\hat{S}}\left(0\right)=\int_{-\infty}^{0}h_{\Omega}(0,t)\hat{\mathcal{E}}_{\text{in}}\left(t\right)dt+\hat{\mathcal{W}}_{\mathcal{S}}(0).\label{eq:S storage alkali}
\end{equation}
The transfer function $h_{\Omega}(0,t)$ then satisfies \begin{equation}
\int_{-\infty}^{0}e^{-2\gamma_{\text{s}}t}|h_{\Omega}(0,t)|^{2}dt\leq\sqrt{\frac{C+1}{C}}.\label{eq:Equality of f}
\end{equation}Maximal storage efficiency is realized by shaping
the temporal profile of the control field $\Omega(t)$ to satisfy
$h_{\Omega}(0,t)=A_{\Omega}\mathcal{\hat{E}}^*_{\text{in}}\left(t\right)$, where the normalization constant $A_{\Omega}$ is given by\begin{equation}
A_{\Omega}=\sqrt{\frac{\int_{-\infty}^0|h_{\Omega}(0,s)|^{2}ds}{\int_{-\infty}^0\langle\mathcal{E}_{\text{in}}^{\dagger}\mathcal{E}_{\text{in}}\rangle_{s}ds}}.\end{equation}
Using Eq.~(\ref{eq:S_vector solution}), we can describe the storage efficiency for any input signal by
\begin{equation}
\eta_{\text{in}}^{(\mathcal{E}\rightarrow\mathcal{S})}=\frac{C}{C+1}\int_{-\infty}^{0}  \langle\hat{\mathcal{E}}_{\text{in}}^\dagger\hat{\mathcal{E}}_{\text{in}}\rangle_t e^{2\gamma_{\textrm{s}}t} dt,\label{eq:3-level-efficiency}
\end{equation}
which approaches $C/(C+1)$ in the short pulse limit ($\gamma_{\textnormal{s}}T\ll1$), but otherwise depends on the temporal mode function of the input field.

\smallskip{}

\emph{Retrieval}.--- The output field during retrieval, obtained by substituting Eq.~(\ref{eq:P adiabatic-1}) in Eq.~(\ref{eq:output-light-P}), is
\begin{equation}
\hat{\mathcal{E}}_{\text{out}}\left(t\right)=\alpha\hat{\mathcal{E}}_{\text{in}}-Q\Omega\hat{\mathcal{S}}+\hat{f}_{\mathcal{E}},\label{eq:output light spins single mode}\end{equation} where $\alpha$ is given in Eq.~(\ref{eq:alpha_definition}). Note that Eq.~(\ref{eq:output light spins single mode}) is the single-mode version of Eq.~(\ref{eq:output light spins}).
The output field squared for any $t\geq\tau$ is then given by
\begin{align}
\langle\hat{\mathcal{E}}_{\text{out}}^\dagger\hat{\mathcal{E}}_{\text{out}}\rangle_{t} & =|h_{\Omega}(t,\tau)|^2\langle\hat{\mathcal{S}}^\dagger\hat{\mathcal{S}}\rangle_{\tau},\label{eq:Alkali retrieval}
\end{align}
assuming that $\hat{f}_{\mathcal{E}}$ satisfies vacuum properties and that  $\langle\hat{\mathcal{E}}_{\text{in}}^\dagger\hat{\mathcal{E}}_{\text{in}} \rangle_{(t>0)}=0$. 
The retrieval efficiency into some target temporal mode $f(t)$ is given by 
\begin{equation}
\eta_{\text{out}}^{(\mathcal{S}\rightarrow\mathcal{E})}=\frac{C}{C+1}\frac{1}{\int_{\tau}^{\infty}|f(t)|^2e^{2\gamma_{\text{s}}(t-\tau)}dt},\label{eq:retreival efficiency alkali}
\end{equation}where $f(t)$ is normalized such that $\int_{\tau}^{\infty}|f(t)|^2dt=1$. Like the storage efficiency, the retrieval efficiency approaches  $C/(C+1)$ in the short pulse limit ($\gamma_{\textnormal{s}}T\ll1$), but otherwise depends on the desired temporal mode function of the output field.

\section{Multi-mode description of the sequential mapping\label{sec:multi-mode-sequential}}

In this appendix, we derive the solution for the multi-mode exchange
evolution in the second stage $\hat{\mathcal{S}}\rightarrow\hat{\mathcal{K}}$
of the sequential mapping scheme. Using the diffusion eigenmodes for
the alkali spins in Eq.~(\ref{eq:alkali-diffusion-equation}), the
dynamics is described by
\begin{align}
\partial_{t}\hat{\mathcal{S}}_{m} & =-(\gamma_{\text{s}}^{\text{\ensuremath{\left(m\right)}}}+i\delta_{\text{s}})\hat{\mathcal{S}}_{m}-iJ\sum_{n}c_{mn}\hat{\mathcal{K}}_{n}+\hat{f}_{\mathcal{S}}^{(m)},\label{eq:S_m mode dark-1-2}\\
\partial_{t}\hat{\mathcal{K}}_{n} & =-(\gamma_{\text{k}}^{\text{\ensuremath{\left(n\right)}}}+i\delta_{\text{k}})\mathcal{K}_{n}-iJ\sum_{m}c_{mn}^{*}\mathcal{S}_{m}+\hat{f}_{\mathcal{K}}^{(n)}.\label{eq:K_m mode dark-1-2}
\end{align}
The spins experience coherent dynamics in the dark, with the alkali
and noble-gas spin modes periodically exchanging excitations. The
coefficients $c_{mn}$ {[}Eq.~(\ref{eq: c coefficient  diffusion modes}){]}
weigh the coupling of the $m^{\text{th}}$ mode of one spin gas with
the $n^{\text{th}}$ mode of the other spin gas. The solution of Eqs.~(\ref{eq:S_m mode dark-1-2}-\ref{eq:K_m mode dark-1-2})
reads
\begin{equation}
\left(\begin{array}{c}
\boldsymbol{\mathcal{\hat{S}}}(T')\\
\boldsymbol{\mathcal{\hat{K}}}(T')
\end{array}\right)=\Psi_{T',0}\left(\begin{array}{c}
\boldsymbol{\mathcal{\hat{S}}}(0)\\
\boldsymbol{\mathcal{\hat{K}}}(0)
\end{array}\right)+\left(\begin{array}{c}
\boldsymbol{\mathcal{\hat{W}}}_{\text{s}}(T')\\
\boldsymbol{\mathcal{\hat{W}}}_{\text{k}}(T')
\end{array}\right),\label{eq:alkali-noble gas solution}
\end{equation}where the matrix $\Psi_{T',T}$ describes the evolution of the spins
from time $t=0$ to time $T'$, and the vector of stochastic operators
is given by
\begin{equation}
\left(\begin{array}{c}
\boldsymbol{\mathcal{\hat{W}}}_{\text{s}}(T')\\
\boldsymbol{\mathcal{\hat{W}}}_{\text{k}}(T')
\end{array}\right)=\int_{0}^{T'}\Psi_{T',t}\left(\begin{array}{c}
\boldsymbol{\hat{f}}_{\mathcal{S}}\left(t\right)\\
\boldsymbol{\hat{f}}_{\mathcal{K}}\left(t\right)
\end{array}\right)dt.\label{eq:stochastic diffusion operators}
\end{equation}
For a constant magnetic field during the interaction, $\Psi_{T',0}$
is given by
\begin{equation}
\Psi_{T',0}=\exp\left[\left(\begin{array}{cc}
[\text{A}_{\text{s}}] & iJ\left[c\right]\\
iJ\left[c\right]^{\dagger} & [\text{A}_{\text{k}}]
\end{array}\right)T'\right],
\end{equation}
where the matrices $[\text{A}_{\text{s}}]$, $[\text{A}_{\text{k}}]$,
and $[c]$ have the elements $[\text{A}_{\text{s}}]_{mn}=(\gamma_{\text{s}}^{(m)}+i\delta_{\text{s}})\delta_{mn}$,
$[\text{A}_{\text{k}}]_{mn}=(\gamma_{\text{k}}^{(n)}+i\delta_{\text{k}})\delta_{mn}$,
and $[c]_{mn}=c_{mn}$.

The exchange evolution depends on the detuning $\text{\ensuremath{\delta}}$
between the alkali and noble-gas spins, and the coupling is maximal
on resonance $\delta=0$. If the quantum signal is mapped on the uniform
mode of the alkali spins at storage, then after a $\pi$ pulse {[}with
$T'\approx\pi/(2J)${]}, we find the efficiency
\begin{equation}
\eta_{\text{in}}^{\left(\mathcal{S}\rightarrow\mathcal{K}\right)}=\frac{\langle\hat{\mathcal{K}}^\dagger\hat{\mathcal{K}}\rangle_{T'}}{\langle\hat{\mathcal{S}}^\dagger\hat{\mathcal{S}}\rangle_{\left(t=0\right)}}=\sum_{m}|c_{0m}|^{2}\exp\Bigl(-\frac{\gamma_{m}^{\text{(s)}}\pi}{2J}\Bigr).\label{eq:sequential_Storage_efficiency_Step2}
\end{equation}

\end{document}